\pdfoutput=1
\documentclass[final]{siamart190516}

\usepackage{epstopdf}
\usepackage{hyperref}
\usepackage{layout}

\usepackage[utf8]{inputenc} 
\usepackage[T1]{fontenc}    
\usepackage{hyperref}       
\usepackage{url}            
\usepackage{booktabs}       
 
\usepackage{nicefrac}       
\usepackage{microtype}      
\usepackage{lipsum}		
\usepackage{graphicx}
\usepackage[numbers]{natbib}
\usepackage{doi}

\usepackage{mathtools}
\usepackage{amsmath,amssymb, amsfonts}

\usepackage{color, colortbl}
\definecolor{Gray}{gray}{0.95}
\definecolor{lightgray}{gray}{0.95}

\usepackage{etoolbox} 
\usepackage{lipsum} 

\usepackage[first=0,last=9]{lcg}

\newcommand{\bvec}[1]{\mbox{\boldmath $#1$}}
\newcommand{\R}{\mathbb{R}}

\newcommand*{\ditto}{---\texttt{"}---}

\usepackage{algorithm,algorithmic,amssymb,amsmath,letltxmacro}
\newlength{\commentindent}
\makeatletter
\renewcommand{\algorithmiccomment}[1]{\unskip\hfill\makebox[\commentindent][l]{//~#1}\par}
\LetLtxMacro{\oldalgorithmic}{\algorithmic}
\renewcommand{\algorithmic}[1][0]{%
  \oldalgorithmic[#1]%
  \renewcommand{\ALC@com}[1]{%
    \ifnum\pdfstrcmp{##1}{default}=0\else\algorithmiccomment{##1}\fi}%
}
\makeatother

\newsiamremark{remark}{Remark}
\newsiamremark{hypothesis}{Hypothesis}
\crefname{hypothesis}{Hypothesis}{Hypotheses}
\newsiamthm{claim}{Claim}

\ifpdf
  \DeclareGraphicsExtensions{.eps,.pdf,.png,.jpg}
\else
  \DeclareGraphicsExtensions{.eps}
\fi

\usepackage{amsopn}

\title{Communication-hiding pipelined BiCGSafe methods for solving large linear systems}

\author{Viet Q. H. HUYNH $^*$, Hiroshi SUITO \thanks{Advanced Institute for Materials Research, Tohoku University,
 2-1-1 Katahira, Aoba-ku, Sendai, 980-8577, Japan. Correspondent: hqhviet@tohoku.ac.jp}}

\makeatletter
\newcommand*{\addFileDependency}[1]{
  \typeout{(#1)}
  \@addtofilelist{#1}
  \IfFileExists{#1}{}{\typeout{No file #1.}}
}
\makeatother

\begin{document}
\maketitle
\begin{abstract}

Recently, a new variant of the BiCGStab method, known as the pipelined BiCGStab, has been proposed. This method can achieve a higher degree of scalability and speed-up rates through a mechanism in which the communication phase for the computation of the inner product can be overlapped with computation of the matrix-vector product. Meanwhile, several generalized iteration methods with better convergence behavior than BiCGStab exist, such as ssBiCGSafe, BiCGSafe, and GPBi-CG. Among these methods, ssBiCGSafe, which requires a single phase of computing inner products per iteration, is best suited for high-performance computing systems. As described herein, inspired by the success of the pipelined BiCGStab method, we propose pipelined variations of the ssBiCGSafe method in which only one phase of inner product computation per iteration is required and this phase of inner product computation can be overlapped with the matrix-vector computation. Through numerical experimentation, we demonstrate that the proposed methods engender improvements in convergence behavior and execution time compared to the pipelined BiCGStab and ssBiCGSafe methods.

\end{abstract}


\section{Introduction}

\begin{figure}[htbp]
\caption{Global reduction scheme of the inner product on distributed memory HPC systems
}
\label{fig:dot_product}
\centering
\includegraphics[width=12cm]{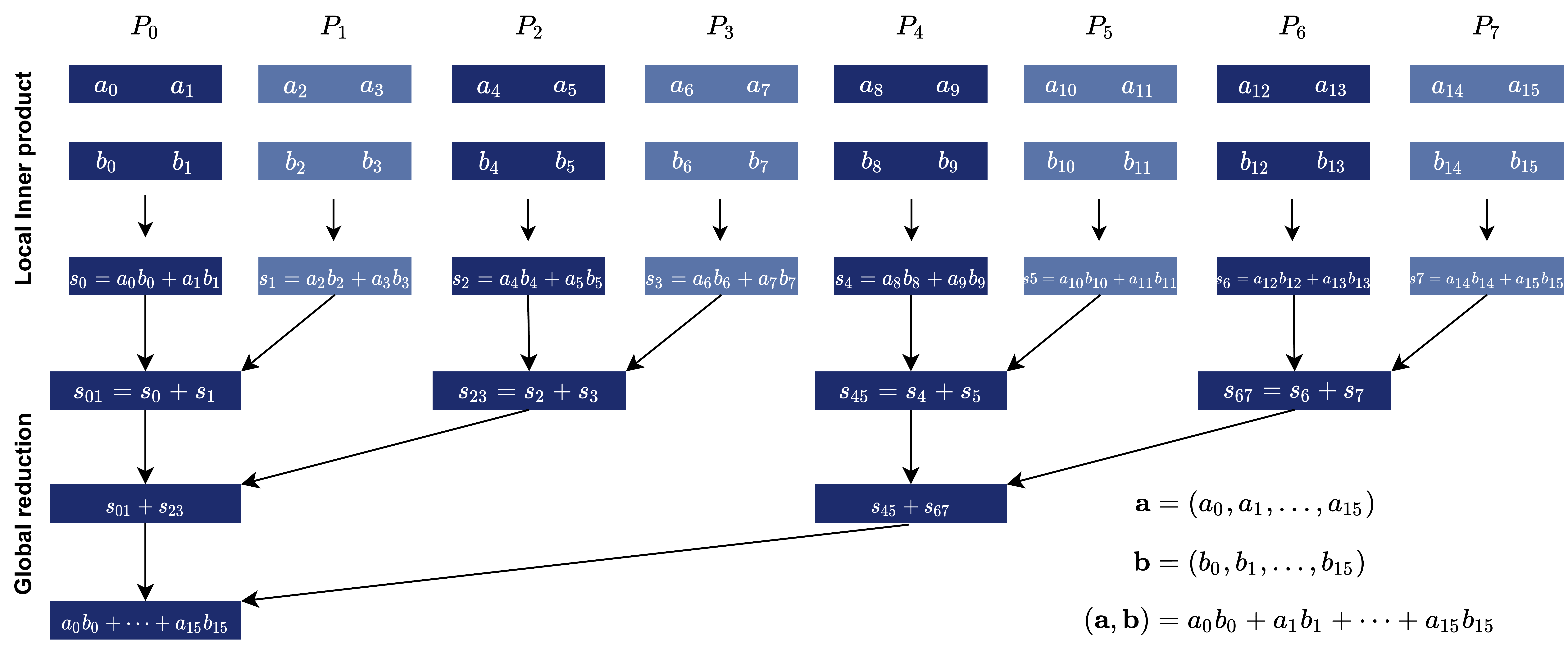}
\end{figure}

Many scientific and engineering applications lead to problems of solving systems of linear equations $\bvec{Ax}=\bvec{b}$ for an unknown vector $\bvec x  \in \R^n$, where $\bvec{A} \in \R^{n\times n} $ is a given sparse matrix that is typically very large, and where $\bvec b \in \R^n$ is a given dense right-hand side vector. One method often used to solve systems of linear equations is the iterative BiCGStab algorithm \cite{vorst1992}, which consists of three basic operations: an inner product, linear combination, and matrix-vector product. Initialized from an arbitrary solution, the best approximate solution can be derived from a sequence of improved approximate solutions. Real-life applications necessitate the acceleration of the solving process through the use of high-performance computing (HPC) systems to obtain a solution within a certain period of time. When executing the BiCGStab algorithm on HPC systems with distributed memory, the inner product computation is the main process that engenders delays in the entire system. The inner product computation requires a global reduction operation to collect scalar subtotals in each processor to one processor, as shown in Fig. \ref{fig:dot_product} and requires a global communication operation to distribute the result to all processors. Time for inner product computation is expected to dominate the time of the whole algorithm as the number of processors increases.\\

Based on the BiCGStab method, several generalized algorithms with better convergence behavior have been developed further and proposed, such as ssBiCGSafe \cite{Fujino2015}, BiCGSafe \cite{Fujino2005}, GPBi-CG \cite{Zhang1997}, GPBiCG v1..,v4 \cite{Abe2013}. Among these methods, which typically need two global reduction phases for computation on inner products per iteration, the ssBiCGSafe method, which requires only one global reduction phase, was found to be highly efficient in terms of the convergence rate and elapsed time through many numerical experiments on HPC systems with distributed memory \cite{Fujino2015,Fujjino2018}.

The increase in speedup rates corresponding to the increase in computing nodes, also known as scalability, is a criterion for evaluation of the parallel algorithm effectiveness. In HPC systems with numerous compute nodes, scalability is highly dependent on the communication time between compute nodes. One method to reduce communication costs is to hide them through other computation processes. As a method to increase BiCGStab scalability, Cools and Vanroose have proposed a new variant of the BiCGStab method by which communication latency of inner product computations can be hidden in a matrix-vector computation \cite{Cools}. On distributed memory HPC systems, the pipelined BiCGStab method (abbreviated as p-BiCGStab) can gain a higher scalability property than that of the standard BiCGStab method. As described in this paper, similar to the work of Cools and Vanroose, we propose variants of the ssBiCGSafe method with better convergence behavior than that of the pipelined BiCGStab method, and better speedup rates on HPC systems than that of the standard ssBiCGSafe method.

This paper is organized as follows. Section 2 presents an overview of the ssBiCGsafe method. Section 3 explains derivation of the new pipelined BiCGSafe method designated as p-BiCGSafe. For applications which require high accuracy, the proposed p-BiCGSafe method can be adversely affected by deviation of the residual vector, which causes it to stagnate at a value higher than the required accuracy. Section 4 presents a variant of the p-BiCGsafe method, named p-BiCGSafe-rr, which addresses this situation by employing the residual replacement technique. Section 5 shows numerical experiments that were conducted to verify the convergence behavior and speedup rates of the proposed methods. The last section provides a brief summary of the proposed methods.

\section{Overview of the BiCGSafe Method with Single Reduction Phase}
Given a matrix of real numbers $\bvec A \in \R ^{n \times n}$ and a right-hand side vector $\bvec b \in \R^n$, one can consider the linear system of equations $\bvec A\bvec x = \bvec b$, where $\bvec x \in \R^n$  is an unknown vector. $\bvec x^* $ denotes a solution vector. Most iterative methods start with an initial solution $ \bvec x_0$ and create a series of vectors $\bvec x_n $ that approximate solution vector $\bvec x^*$.
Letting $\bvec r_0 = \bvec b - \bvec A \bvec x_0$ be the initial residual vector, then the Krylov subspace of order $i$ constructed by the matrix $\bvec A$ and the initial residual vector $\bvec r_0$ is the linear subspace spanned by the images of vector $\bvec r_0$ under the first $i$ powers of the matrix $\bvec A$ as \\
${\mathcal {K}}_{i}(\bvec A,\bvec r_0)=\operatorname {span} \,\{\bvec r_0,\bvec A \bvec r_0 ,\bvec A^{2} \bvec r_0,\ldots ,\bvec A^{i-1} \bvec r_0\}$.
Here, $\mathcal {K}_i (\bvec A, \bvec r_0)$ is the set of all linear combinations as
$a_0\bvec r_0 + a_1\bvec A \bvec r + a_2\bvec A^2\bvec r_0 + \cdots + a_{i-1}\bvec A^{i-1} \bvec r_0$. In the Krylov subspace method, the approximate solution of the linear system can be expressed as
$\bvec x_i = \bvec x_0 + \bvec z_i$, where $\bvec z_i \in \mathcal {K}_i(\bvec A, \bvec r_0)$. The corresponding $ i^{th}$ residual vector can be written as
$\bvec r_i = \bvec b - \bvec A \bvec x_i = \bvec r_0 - \bvec A \bvec z_i \in \mathcal {K}_{i+1}(\bvec A, \bvec r_0)$. The residual vector can be computed by the product of the polynomial in $\bvec A$ and the initial residual vector $\bvec r_0$.
For BiCG algorithm, the residual vector is definable by a polynomial of $\bvec A$ with degree $i$ as
\begin{align}
\bvec r^{BiCG}_i = R_i(A)\bvec r_0.
\end{align}
Here, $R_i(A)$ is designated as the residual polynomial of BiCG. In fact, it is a multiple of the residual vector with a Bi-Lanczos polynomial that satisfies the following recursive relation as  
\begin{align}\label{eq:BiLanczos}
\begin{cases}
R_0(\bvec A) &= \bvec I, \; \; \; \;  R_1(\bvec A) = (\bvec I - \alpha_0 \bvec A) R_0 (\bvec A),\\
R_{i+1}(\bvec A) &= (\bvec I + \frac{\beta_{i-1}}{\alpha_{i-1}}\alpha_i \bvec I - \alpha_i \bvec A)R_i(\bvec A)-  \frac{\beta_{i-1}}{\alpha_{i-1}}\alpha_i R_{i-1}(\bvec A),  \; \; i = 1,2,\cdots 
\end{cases}
\end{align}

for certain coefficients $\alpha_{i}$ and $\beta_{i-1}$.

In the BiCGStab algorithm, the residual is defined as

\begin{align} \label{eq:BiCGStabResidual}
\bvec r^{BiCGStab}_i &= P_i(\bvec A)\bvec r^{BiCG}\\
& = P_i(\bvec A)R_i(\bvec A)r_0,
\end{align}

where $P_i(A)$, called a stabilizing polynomial, is a polynomial of degree $i$, defined recursively as 

\begin{align}\label{eq:BiCGStabStabilizingPolynomial}
\begin{cases}
P_0(\bvec A) &= \bvec I,\\
P_i(\bvec A) &= (\bvec I - \omega_{i-1} \bvec A)P_{i-1}(\bvec A), \; \; i = 1,2,\cdots
\end{cases}
\end{align}

Here, $I$ represents the unit matrix. Also, $\omega_i $ is a nonzero coefficient chosen to provide stable convergence behavior. 


\begin{algorithm}[H]
\caption{BiCGStab}
\label{alg:BiCGStab}
\begin{algorithmic}[1]
\STATE Let $\bvec{x}_0$ be an initial guess;
\STATE Compute $\bvec{r}_0 = \bvec{b}- \bvec{Ax}_0$, and set $\beta_{-1} = 0 $;
\STATE Choose $\bvec{r}_0^*$ such that $(\bvec{r}_0^*,\bvec{r}_{0}) \neq 0$, e.g., $\bvec{r}_0^* = \bvec{r}_{0}$;
\FOR {$n=0,1...$}

\STATE \textbf{if} $||\bvec{r}_{i}||/||\bvec{r}_0||\leq \epsilon $ \textbf{stop};
\STATE $\bvec p_{i} = \bvec r_{i} + \beta_{i-1} ( \bvec p_{i-1} -\zeta_{i-1} \bvec A \bvec p_{i-1} ) $;
\STATE Compute $\bvec{Ap}_{i}$;
\STATE $\alpha_{i} = \frac{(\bvec{r}_0^*,\bvec{r}_{i})}{ (\bvec r_0^*, \bvec{A}\bvec{p}_{i})}$;

\STATE $\bvec{t}_{i} = \bvec{r}_{i} - \alpha_{i} \bvec{Ap}_{i} $;
\STATE Compute $\bvec{At}_{i}$;  

\STATE $\omega_{i} = \frac{(\bvec{At}_{i},\bvec{t}_{i})}{ (\bvec{At}_{i},\bvec{At}_{i})}$;

\STATE	$\bvec{x}_{i+1} = \bvec{x}_{i} +\alpha_{i}\bvec{p}_{i} + \zeta_{i} \bvec{t}_{i}$;

\STATE	$\bvec{r}_{i+1} = \bvec{t}_{i} - \zeta_{i} \bvec{At}_{n}$;
\STATE $\beta_{i} = \frac{\alpha_{i}}{\omega_{i} } \frac{(\bvec{r}_0^*,\bvec{r}_{i+1})}{ (\bvec r_0^*, \bvec{r}_{i})}$;

\ENDFOR
\end{algorithmic}
\end{algorithm}

To obtain faster and more stable convergence properties, Zhang \cite{Zhang1997}  proposed a generalization of stabilizing polynomials constructed in the form of three-term recurrences as of Lanczos polynomials in (\ref{eq:BiLanczos}) by adding two independent coefficients $\zeta_i$ and $\eta_i$ as shown below.

\begin{align} \label{eq:7}
\begin{cases}
&P_0(\bvec A) = \bvec I,\; \; \; \; P_1(A) = (\bvec I - \zeta_0\bvec A) P_0 (\bvec A),\\
&P_{i+1}(\bvec A) = (\bvec I+\eta_i\bvec I -\zeta_i \bvec A)P_i(\bvec A) -\eta_i P_{i-1}(\bvec A), \; \; i = 1,2,\cdots
\end{cases}
\end{align}

It is noteworthy that setting $\eta_i = 0$ and $\zeta_i = \omega_i $ in these polynomials yields the form of the stabilizing polynomial (\ref{eq:BiCGStabStabilizingPolynomial}) of the BiCGStab algorithm. In the actual construction of the polynomial $P_i$, an auxiliary polynomial $G_i$ of degree $i$,

\begin{align} \label{eq:8}
G_{i}(\bvec A) = \zeta_i^{-1}\bvec A^{-1} (P_i(\bvec A) - P_{i+1}(\bvec A) ),
\end{align}

 was used to reconstruct the three-term recurrences (\ref{eq:7}) into the following interlaced two-term recurrences. 

\begin{align} \label{eq:9}
\begin{cases}
&P_{0}(\bvec A) = \bvec I,\; \; \; \; G_0(\bvec A) = \bvec I,\\
&P_{i+1}(\bvec A) = P_{i}(\bvec A) - \zeta_i \bvec A G_i(\bvec A),\\
&G_{i+1}(\bvec A) = P_{i+1}(\bvec A) + \zeta_i \frac{\eta_{i+1}}{\zeta_{i+1}} G_{i}(\bvec A), \; \; i = 1,2,\cdots
\end{cases}
\end{align}

Based on this generalization of the stabilizing polynomials, Zhang proposed a generalized product type method name GPBi-CG based on the BiCG method for solving the linear system, as shown in Algorithm \ref{alg:GPBiCG} \cite{Zhang1997}. Here, the residual is constructed as shown in line 24:
$\bvec r_{i+1} = \bvec t_i - \zeta_i\bvec y_i -\eta_i \bvec {At}_i$; and the recurrence coefficients $\zeta_i$ and $\eta_i$ are computed such that the residual norm $\|\bvec r_{i+1}\|_2 = \| \bvec t_i - \zeta_i\bvec y_i -\eta_i \bvec {At}_i \|_2$ is minimized. This process is equivalent to finding the coefficients $\zeta_i$ and $\eta_i$ to satisfy the following orthogonal relationships:

\begin{align}
\begin{cases}
\bvec t_i - \zeta_i\bvec y_i -\eta_i \bvec {At}_i \perp \bvec y_i,  \\
\bvec t_i - \zeta_i\bvec y_i -\eta_i \bvec {At}_i \perp \bvec {At}_i. 
\end{cases}
\end{align}

Consequently, the values of $\eta_i$ and $\zeta_i$ are the values that solve the next $2 \times 2$ linear system:

\begin{align} \label{eq:eta&zeta}
\begin{cases}
 (\bvec y_i, \bvec y_i) \eta_i + (\bvec y_i, \bvec {At}_i)\zeta_i &= (\bvec y_i, \bvec t_i),\\
 (\bvec {At}_i, \bvec {y}_i) \eta_i + (\bvec {At}_i, \bvec {At}_i)\zeta_i &= (\bvec {At}_i, \bvec t_i).
\end{cases}
\end{align}

The values of $\eta_i$ and $\zeta_i$ that solves the linear system (\ref{eq:eta&zeta}) is given at lines 18 and 19 in Alg. \ref{alg:GPBiCG}.

Coefficients $\alpha_i = (\bvec{r}_0^*,\bvec{r}_i) / (\bvec r_0^*, \bvec{A}\bvec{p}_i)$ and $\beta_i = \alpha_i/\zeta_i  \times (\bvec{r}_0^*,\bvec{r}_{n+1}) / (\bvec r_0^*, \bvec{r}_i)$ are mathematically equivalent to coefficients $\alpha_i$ and $\beta_i$ of the BiCG method (see \cite{Zhang1997}). It is noteworthy that if one substitutes $\eta_i=0$ and $\zeta_i=\omega_i = (\bvec {At}_i,\bvec t_i)/(\bvec {At}_i,\bvec {At}_i$) for any $i$ in Algorithm \ref{alg:GPBiCG}, one obtains the BiCGStab algorithm \ref{alg:BiCGStab}. Actually, GPBi-CG is useful as a more effective alternative to BiCGStab in numerical simulations \cite{Surabhi2021, Viet2017}.

\begin{algorithm}[H]
\caption{GPBi-CG}
\label{alg:GPBiCG}
\begin{algorithmic}[1]

\STATE Let $\bvec{x}_0$ be an initial guess; 
\STATE Compute $\bvec{r}_0 = \bvec{b}- \bvec{Ax}_0$; 
\STATE Choose $\bvec{r}_0^*$ such that $(\bvec{r}_0^*,\bvec{r}_{0}) \neq 0$, e.g., $\bvec{r}_0^* = \bvec{r}_{0}$;
\STATE Set $\bvec p_{-1} =  \bvec u_{-1} = \bvec t_{-1} = \bvec w_{-1} = \bvec 0$, and $\beta_{-1} = 0$;

\FOR {$i=0,1...$}
\STATE \textbf{if} $||\bvec{r}_i||/||\bvec{r}_0||\leq \epsilon $ \textbf{stop}.
\STATE	$\bvec{p}_i = \bvec{r}_i +\beta_{i-1}(\bvec{p}_{i-1}-\bvec{u}_{i-1})$;
\STATE Compute $\bvec{Ap}_i$; 
\STATE $\alpha_i = \frac{(\bvec{r}_0^*,\bvec{r}_i)}{ (\bvec r_0^*, \bvec{A}\bvec{p}_i)}$;

\STATE $\bvec{y}_{i} = \bvec{t}_{i-1} - \bvec r_i - \alpha_i\bvec{w}_{i-1} + \alpha_i\bvec{Ap}_i $;

\STATE $\bvec{t}_i = \bvec{r}_i -\alpha_i \bvec{Ap}_i $;
\STATE Compute $\bvec{At}_i$; 
\STATE Define $a_i \coloneqq (\bvec{y}_i,\bvec{y}_i)$,
  $b_i \coloneqq (\bvec{At}_i,\bvec{t}_i)$,
  $c_i \coloneqq (\bvec{y}_i,\bvec{t}_i)$,
  $d_i  \coloneqq(\bvec{At}_i,\bvec{y}_i)$,
  $e_i \coloneqq (\bvec{At}_i,\bvec{At}_i)$;

\IF {$i=0$} 
\STATE $\zeta_i=b_i/e_i$;
\STATE $\eta_i=0$; 
\ELSE
\STATE $\zeta_i = (a_ib_i -c_id_i)/(e_ia_i- d_i^2) $;
\STATE $\eta_i  = (e_ic_i -d_ib_i)/(e_ia_i- d_i^2) $;
\ENDIF

\STATE	$\bvec{u}_i = \zeta_i \bvec{Ap}_i +\eta_i (\bvec t_{i-1} - \bvec{r}_i+\beta_{i-1}\bvec{u}_{i-1})$;
\STATE $\bvec{z}_i = \zeta_i\bvec{r}_i + \eta_i\bvec{z}_{i-1}-\alpha_i\bvec{u}_i $;
\STATE $\bvec{x}_{i+1} = \bvec{x}_i+\alpha_i \bvec{p}_i + \bvec{z}_i $;
\STATE $\bvec{r}_{i+1} = \bvec{t}_i - \eta_i \bvec{y}_i -\zeta_i \bvec{At}_i$;

\STATE $\beta_i = \frac{\alpha_i}{\zeta_i } \frac{(\bvec{r}_0^*,\bvec{r}_{n+1})}{ (\bvec r_0^*, \bvec{r}_i)}$;

\STATE $\bvec{w}_i = \bvec{At}_i  + \beta_i\bvec{Ap}_i$;
\ENDFOR
\end{algorithmic}
\end{algorithm}

As shown at lines 9, 14--20, and 25 of Alg. \ref{alg:GPBiCG}, it requires three iterations of synchronization per iteration to compute the inner products. In an earlier work \cite{Fujino2005}, Fujino proposed a variant of the GPBi-CG named BiCGSafe with a better convergence property based on a new strategy of constructing an associate residual. This strategy reduces the number of synchronizations from three to two per iteration. Fujino also proposed improved variants of the BiCGSafe with single synchronization: ssBiCGSafe1 and ssBiCGSafe2 \cite{Fujino2015}. The salient difference between the two is the use of a transposed matrix: ssBiCGSafe1 uses it, whereas ssBiCGSafe2 does not. In HPC systems, additional processing steps must be followed to compute the transpose matrix. The version of BiCGSafe method without using a transpose matrix has benefits. We specifically examine derivation of the pipelined version of the ssBiCGSafe2 method: Alg. \ref{alg:ssBiCGSafe2} shows ssBiCGSafe2, which is presented at \cite{Fujino2013}. Some variable names are modified in the next section for convenient derivation of pipelined versions. We also change the position of checking the convergence so that computation of the inner product of the residual vector with itself can be performed simultaneously with that of the other inner products.

In GPBi-CG, residual vector $\bvec r_{i}$ is computed based on coefficients $\eta_i$ and $\zeta_i$, which are selected such that the norm of residual vector $\bvec{r}_i$ is minimized. However, in BiCGSafe methods, the residual vector does not include coefficients $\eta_i$ and $\zeta_i$, as shown in line 30 of Alg. \ref{alg:ssBiCGSafe2}. The following associate residual vector $\bvec {a\_r}_i$ is defined as

\begin{align}
\bvec{a\_r}_i \coloneqq \bvec{r}_i -\zeta_i \bvec {Ar}_i - \eta \bvec {y}_i.
\end{align}

\begin{algorithm}[H]
\caption{ssBiCGSafe2}
\label{alg:ssBiCGSafe2}
\begin{algorithmic}[1]
\STATE Let $\bvec{x}_0$ be an initial guess; 
\STATE Compute $\bvec{r}_0 \gets \bvec{b}- \bvec{Ax}_0$;
\STATE Choose $\bvec{r}_0^*$ such that $(\bvec{r}_0^*,\bvec{r}_{0}) \neq 0$, e.g., $\bvec{r}_0^* = \bvec{r}_{0}$;

\FOR {$\textit{i = 0, 1...}$}
\STATE Compute $\bvec s_i \gets \bvec{Ar}_i$; \COMMENT{matrix-vector product}
\STATE Define $a_i \coloneqq (\bvec{s}_i,\bvec{s}_{i})$,
  $b_i \coloneqq (\bvec{y}_i,\bvec{y}_{i})$,
  $c_i \coloneqq (\bvec{s}_i,\bvec{y}_{i})$,
  $d_i \coloneqq (\bvec{s}_i,\bvec{r}_{i})$;
\STATE Define $e_i \coloneqq (\bvec{y}_i,\bvec{r}_{i})$,
  $f_i \coloneqq (\bvec{r}_0^*,\bvec{r}_{i})$,
  $g_i \coloneqq (\bvec{r}_0^*,\bvec{s}_{i})$,
  $h_i \coloneqq  (\bvec{r}_0^*,\bvec{t}_{i-1})$,
  $r_i \coloneqq  (\bvec{r}_i,\bvec{r}_i)$;

\IF {$i=0$} 
\STATE Compute $a_i, d_i, f_i, g_i,r_i$; \COMMENT{inner products}
\STATE $\beta_i \gets 0$; 
\STATE $\alpha_i \gets f_i/g_i$;
\STATE $\zeta_i \gets d_i/a_i$;
\STATE $\eta_i \gets 0$; 
\ELSE
\STATE Compute $a_i,b_i,c_i, d_i,e_i, f_i, g_i,h_i,r_i$; \COMMENT{inner products}
\STATE $\beta_i  \gets (\alpha_{i-1}f_i)/(\zeta_{i-1}f_{i-1})$;
\STATE $\alpha_i \gets f_i/(g_i + \beta_i h_i )$;
\STATE $\zeta_i  \gets (b_id_i -c_ie_i)/(a_ib_i- c_i^2) $;
\STATE $\eta_i   \gets (a_ie_i -c_id_i)/(a_ib_i- c_i^2) $;
\ENDIF
\STATE \textbf{if} $||\bvec{r}_i||\leq \epsilon||\bvec{r}_0|| $ \textbf{stop}.\COMMENT{$||\bvec{r}_i||_2=\sqrt{r_i}$}

\STATE	$\bvec{p}_i \gets \bvec{r}_i +\beta_{i}(\bvec{p}_{i-1}-\bvec{u}_{i-1})$;
\STATE $\bvec{o}_i  \gets \bvec{s}_i + \beta_{i}\bvec{t}_{i-1}$;
\STATE $\bvec{u}_i  \gets \zeta_i \bvec{o}_i + \eta_i(\bvec{y}_{i} + \beta_{i}\bvec{u}_{i-1}) $;
\STATE Compute $\bvec w_i \gets \bvec{Au}_i$, \COMMENT{matrix-vector product}
\STATE $\bvec{t}_i \gets \bvec{o}_i - \bvec{w}_i $;
\STATE $\bvec{z}_i \gets \zeta_i\bvec{r}_i + \eta_i\bvec{z}_{i-1}-\alpha_i\bvec{u}_i $;

\STATE $\bvec{y}_{i+1} \gets \zeta_i\bvec{s}_i + \eta_i\bvec{y}_{i} - \alpha_i\bvec{w}_i $;
\STATE $\bvec{x}_{i+1} \gets \bvec{x}_{i}+\alpha_i \bvec{p}_i + \bvec{z}_i $;
\STATE $\bvec{r}_{i+1} \gets \bvec{r}_{i} - \alpha_i \bvec{o}_i -\bvec{y}_{i+1}$;
\ENDFOR
\end{algorithmic}
\end{algorithm}

This associated residual vector is not used directly in the iterative loop but only to build coefficients $\eta_i$ and $\zeta_i$.
The values of $\zeta_i$ and $\eta_i$ shown in Alg. \ref{alg:ssBiCGSafe2} at lines 18 and 19 are fundamentally the values which minimize the norm of the associate vector $\bvec{a\_r}_i$,or, equivalently, solve the following $2 \times 2$ linear system:

\begin{align} \label{eq:eta&zeta_ssBiCGSafe2}
\begin{cases}
 (\bvec y_i, \bvec {Ar}_i) \zeta_i + (\bvec y_i, \bvec {y}_i)\eta_i &= (\bvec y_i, \bvec r_i),\\
 (\bvec {Ar}_i, \bvec {Ar}_i) \zeta_i + (\bvec {Ar}_i, \bvec {y}_i)\eta_i &= (\bvec {Ar}_i, \bvec r_i).
\end{cases}
\end{align}

In parallel implementations, global reduction can be conducted in the computation of the inner products shown at line 16 of Alg. \ref{alg:ssBiCGSafe2}. The values of $\alpha_i$, $\beta_i$, $\eta_i$ and $\zeta_i$ can be evaluated immediately after the stage of computation of the inner products $a_i,b_i...$ and $r_i$.

\section{Derivation of Pipelined BiCGSafe Method}

At lines $9$ and $15$ of the Algorithm \ref{alg:ssBiCGSafe2}, inner products can be computed through a global reduction process on distributed memory computing systems. Except for the first iteration, eight local partial products can be computed from each node and eight results can be collected at one node through a reduction phase. It is particularly notable that at each iteration, to compute the inner products, we must first compute a vector-matrix multiplication (abbreviated as {\it MV}), as shown in line 6.

From Alg. \ref{alg:ssBiCGSafe2}, we develop a variation by which the reduction phase for inner product computation can be conducted as overlapping with the computation of the matrix-vector multiplication. For this reason, the communication costs of the reduction phase can be concealed or hidden in the cost of {\it MV} computation.

Using the recurrence for $\bvec r_{i+1}$ in line $30$,

\begin{align} \label{eq:r}
\bvec r_{i+1} = \bvec r_i -\alpha_i \bvec {o}_i -\bvec y_{i+1},
\end{align}

the {\it MV} computation $\bvec s_i \coloneqq \bvec {Ar}_i$ in line $5$ of Alg. \ref{alg:ssBiCGSafe2} can be rewritten as

\begin{align} \label{eq:s}
\bvec s_{i} \coloneqq \bvec {Ar}_i  &= \bvec {Ar}_{i-1} -\alpha_{i-1} \bvec {Ao}_{i-1} -\bvec {Ay}_{i} \nonumber \\
                            &= \bvec s_{i-1} -\alpha_{i-1} \bvec {q}_{i-1} -\bvec {g}_{i}.
\end{align}

Here we defined two new auxiliary vector variables $\bvec q_i \coloneqq \bvec {Ao}_i$ and $\bvec g_i \coloneqq \bvec {Ay}_i$, and substituted $\bvec {Ar}_{i-1}$ by $\bvec s_{i-1}$. Because vector variable $\bvec s_{i}$ is calculated at the start of the iteration, it can also be calculated at the end of the previous iteration as

\begin{align} \label{eq:sip1}
\bvec s_{i+1} = \bvec s_{i} -\alpha_{i} \bvec {q}_{i} -\bvec {g}_{i+1}.
\end{align}

By introducing another new vector variable $\bvec l_i \coloneqq \bvec {At}_i$, and using recurrence of the vector variable $\bvec o$ (line $23$)

\begin{align} \label{eq:o}
\bvec {o}_i &= \bvec {s}_i + \beta_i \bvec {t}_{i-1}, 
\end{align}

the vector variable $\bvec q_i$ is calculable as

\begin{align} \label{eq:q}
\bvec q_{i} \coloneqq \bvec {Ao}_i &= \bvec {As}_i + \beta_i \bvec {At}_{i-1} \nonumber\\
&= \bvec {As}_i + \beta_i \bvec {l}_{i-1}.
\end{align}

The {\it MV} computation result $\bvec {As}_i$ is not used in the inner product computation phases (line $9$ and line $15$). Therefore, it can perform in an overlapping manner with the inner product computation phase.  

Using the recurrence of the vector variable $\bvec t_i$ in line $26$ gives

\begin{align} \label{eq:t}
\bvec t_{i} = \bvec {o}_i -\bvec w_{i},
\end{align}

where vector variable $\bvec l_i$ can be computed by the following recurrence as 

\begin{align} \label{eq:l}
\bvec l_{i} \coloneqq \bvec {At}_i &= \bvec {Ao}_i - \bvec {Aw}_{i} \nonumber\\
                           &= \bvec {q}_i - \bvec {Aw}_{i}.
\end{align}

This value of the vector variable $\bvec l_i$ can be computed after obtaining the value of $\bvec q_i$ and $\bvec {Aw}_i$. In Eqn. (\ref{eq:q}) the index $i-i$ shows that the value of $\bvec l_i$ in the earlier iteration will be used to compute vector variable $\bvec q_i$.

\begin{algorithm}[H]
\caption{Pipelined BiCGSafe (p-BiCGSafe)}
\label{alg:PipelinedBiCGSafe}
\begin{algorithmic}[1]
\STATE Let $\bvec{x}_0$ be an initial guess; 
\STATE Compute $\bvec {Ax}_0$; \COMMENT{matrix-vector product}

\STATE $\bvec{r}_0 \gets \bvec{b}- \bvec{Ax}_0$;
\STATE Choose $\bvec{r}_0^*$ such that $(\bvec{r}_0^*,\bvec{r}_{0}) \neq 0$, e.g., $\bvec{r}_0^* = \bvec{r}_{0}$;

\FOR {$\textit{i = 0, 1...}$}

\STATE Compute $\bvec{As}_i$; \COMMENT{matrix-vector product}
\STATE Define $a_i \coloneqq (\bvec{s}_i,\bvec{s}_{i})$,
  $b_i \coloneqq (\bvec{y}_i,\bvec{y}_{i})$,
  $c_i \coloneqq (\bvec{s}_i,\bvec{y}_{i})$,
  $d_i \coloneqq (\bvec{s}_i,\bvec{r}_{i})$;
\STATE Define $e_i \coloneqq (\bvec{y}_i,\bvec{r}_{i})$,
  $f_i \coloneqq (\bvec{r}_0^*,\bvec{r}_{i})$,
  $g_i \coloneqq (\bvec{r}_0^*,\bvec{s}_{i})$,
  $h_i \coloneqq  (\bvec{r}_0^*,\bvec{t}_{i-1})$,
  $r_i \coloneqq  (\bvec{r}_i,\bvec{r}_i)$;
\IF {$i=0$} 
\STATE Compute $a_i, d_i, f_i, g_i, r_i$;\COMMENT{inner products}
\STATE $\beta_i \gets 0$; 
\STATE $\alpha_i \gets f_i/g_i$;
\STATE $\zeta_i \gets d_i/a_i$;
\STATE $\eta_i \gets 0$; 
\ELSE
\STATE Compute $a_i,b_i,c_i, d_i,e_i, f_i, g_i,h_i,r_i$;\COMMENT{inner products}
\STATE $\beta_i \gets (\alpha_{i-1}f_i)/(\zeta_{i-1}f_{i-1})$;
\STATE $\alpha_i \gets f_i/(g_i + \beta_i h_i )$;
\STATE $\zeta_i \gets (b_id_i -c_ie_i)/(a_ib_i- c_i^2) $;
\STATE $\eta_i  \gets (a_ie_i -c_id_i)/(a_ib_i- c_i^2) $;
\ENDIF

\STATE \textbf{if} $||\bvec{r}_i|| \leq \epsilon ||\bvec{r}_0||$ \textbf{stop}. \COMMENT{$||\bvec{r}_i||_2=\sqrt{r_i}$}

\STATE	$\bvec{p}_i \gets \bvec{r}_i +\beta_{i}(\bvec{p}_{i-1}-\bvec{u}_{i-1})$;
\STATE $\bvec{o}_i \gets \bvec{s}_i + \beta_{i}\bvec{t}_{i-1}$;

\STATE $\bvec{u}_i \gets \zeta_i \bvec{o}_i + \eta_i(\bvec{y}_{i} + \beta_{i}\bvec{u}_{i-1}) $;
\STATE $\bvec{q}_i \gets \bvec{As}_i + \beta_{i}\bvec{l}_{i-1}$; \COMMENT{$\bvec q_i \coloneqq \bvec {Ao}_i$ (Eqn. \ref{eq:q})}

\STATE $\bvec{w}_i \gets \zeta_i \bvec{q}_i + \eta_i(\bvec{g}_{i} + \beta_{i}\bvec{w}_{i-1}) $;
  \COMMENT{  $\bvec w_i \coloneqq \bvec {Au}_i$ (Eqn. \ref{eq:w})}

\STATE $\bvec{t}_i \gets \bvec{o}_i - \bvec{w}_i $;

\STATE $\bvec{z}_i \gets \zeta_i\bvec{r}_i + \eta_i\bvec{z}_{i-1}-\alpha_i\bvec{u}_i $;

\STATE $\bvec{y}_{i+1} \gets \zeta_i\bvec{s}_i + \eta_i\bvec{y}_{i} - \alpha_i\bvec{w}_i $;

\STATE $\bvec{x}_{i+1} \gets \bvec{x}_{i}+\alpha_i \bvec{p}_i + \bvec{z}_i $;
\STATE $\bvec{r}_{i+1} \gets \bvec{r}_{i} - \alpha_i \bvec{o}_i -\bvec{y}_{i+1}$;

\STATE Compute $\bvec{Aw}_i$; \COMMENT{matrix-vector product}
\STATE $\bvec{l}_i \gets \bvec{q}_i - \bvec{Aw}_i $; \COMMENT{  $\bvec l_i \coloneqq \bvec {At}_i$ (Eqn. \ref{eq:l}) }
\STATE $\bvec{g}_{i+1} \gets \zeta_i\bvec{As}_i + \eta_i\bvec{g}_{i} - \alpha_i\bvec{Aw}_i $;\COMMENT{  $\bvec g_{i+1} \coloneqq \bvec {Ay}_{i+1}$ (Eqn. \ref{eq:g}) }
\STATE $\bvec{s}_{i+1} \gets \bvec{s}_{i} - \alpha_i \bvec{q}_i -\bvec{g}_{i+1}$;\COMMENT{  $\bvec s_{i+1} \coloneqq \bvec {Ar}_{i+1}$ (Eqn. \ref{eq:sip1}) }
\ENDFOR
\end{algorithmic}
\end{algorithm}

Using the recurrence of the vector variable $\bvec u_i$ in line $24$ of Alg. \ref{alg:ssBiCGSafe2}

\begin{align} \label{eq:u}
\bvec{u}_i = \zeta_i \bvec{o}_i + \eta_i(\bvec{y}_{i} + \beta_{i}\bvec{u}_{i-1}),
\end{align}

the {\it MV} computation $\bvec {w}_i \coloneqq \bvec {Au}_i$ at line $25$ of Alg. \ref{alg:ssBiCGSafe2} can be expressed as

\begin{align} \label{eq:w}
\bvec w_{i} \coloneqq \bvec {Au}_i  &= \zeta_i \bvec {Ao}_{i} + \eta_{i} ( \bvec {Ay}_{i} + \beta_i \bvec {Au}_{i-1}) \nonumber \\
       &= \zeta_i \bvec {q}_{i}  + \eta_{i} ( \bvec {g}_{i} + \beta_i \bvec {w}_{i-1}).
\end{align}

Here, we substituted $\bvec {Au}_{i-1}$ by $\bvec w_{i-1}$ and used two auxiliary vector variables $\bvec q_i \coloneqq \bvec {Ao}_i$ and $\bvec g_i \coloneqq \bvec {Ay}_i$.
The vector variable $\bvec g_{i+1}$ in Eqn. (\ref{eq:s}) can be computed using the recurrence of vector variable $\bvec y_{i+1}$ in line $28$ as

\begin{align} \label{eq:g}
\bvec g_{i+1} \coloneqq \bvec {Ay}_{i+1} &= \zeta \bvec {As}_i + \eta \bvec {Ay}_{i} -\alpha \bvec{Aw}_i \nonumber\\
            &= \zeta \bvec {As}_i + \eta \bvec {g}_{i} -\alpha \bvec{Aw}_i.
\end{align}

This value of vector variable $\bvec g_{i+1}$ can be used to compute the vector variable $\bvec w_i$ in the next iteration, as shown in Eqn. (\ref{eq:w}). Using equations (\ref{eq:q}), (\ref{eq:w}), (\ref{eq:l}), (\ref{eq:g}), and (\ref{eq:sip1}), respectively, for the computation of vector variables $\bvec q_i, \bvec w_i, \bvec l_i, \bvec g_{i+1}, \bvec s_{i+1}$, Alg.  \ref{alg:ssBiCGSafe2} can be rewritten as Alg. \ref{alg:PipelinedBiCGSafe}.

In Alg. \ref{alg:PipelinedBiCGSafe}, $\bvec{As}_i$ (line $6$) is the vector variable that includes the result of the product of matrix $\bvec{A}$ and vector $\bvec {s}_i$. The same notation is also applicable to the vector variable $\bvec{Aw}_i$ (line $33$).
The computation expressions of vector variables $\bvec p_i, \bvec o_i, \bvec u_i$ and $\bvec t_{i}, \bvec z_{i}, \bvec y_{i+1}, \bvec x_{i+1}, \bvec r_{i+1}$ of Alg. \ref{alg:PipelinedBiCGSafe} are identical to those of Alg. \ref{alg:ssBiCGSafe2}. The matrix-vector product $\bvec s_i \gets \bvec{Ar}_i$ of Alg. \ref{alg:ssBiCGSafe2} is replaced by the recurrences  $\bvec s_{i+1} \gets \bvec {s}_{i}  - \alpha_{i} \bvec {q}_{i} - \bvec {g}_{i+1}$ and $\bvec g_{i+1}  \gets \zeta \bvec {As}_i + \eta_i \bvec {g}_{i} -\alpha_i \bvec{Aw}_i$ in Alg. \ref{alg:PipelinedBiCGSafe}. The matrix-vector product $\bvec w_i \gets \bvec{Au}_i$ of Alg. \ref{alg:ssBiCGSafe2} is replaced by the recurrences  $\bvec w_i \gets  \zeta_i \bvec {q}_{i}  + \eta_{i} ( \bvec {g}_{i} + \beta_i \bvec {w}_{i-1})$, here, $\bvec q_i \gets \bvec {As}_i + \beta_i \bvec {l}_{i-1}$, and $\bvec l_i \gets \bvec {q}_i - \bvec {Aw}_{i}$ in Alg. \ref{alg:PipelinedBiCGSafe}.

\begin{figure}[h]
\caption{Computation phases per iteration of iterative methods of four kinds}
\label{fig:diagram}
\centering
\includegraphics[width=\linewidth]{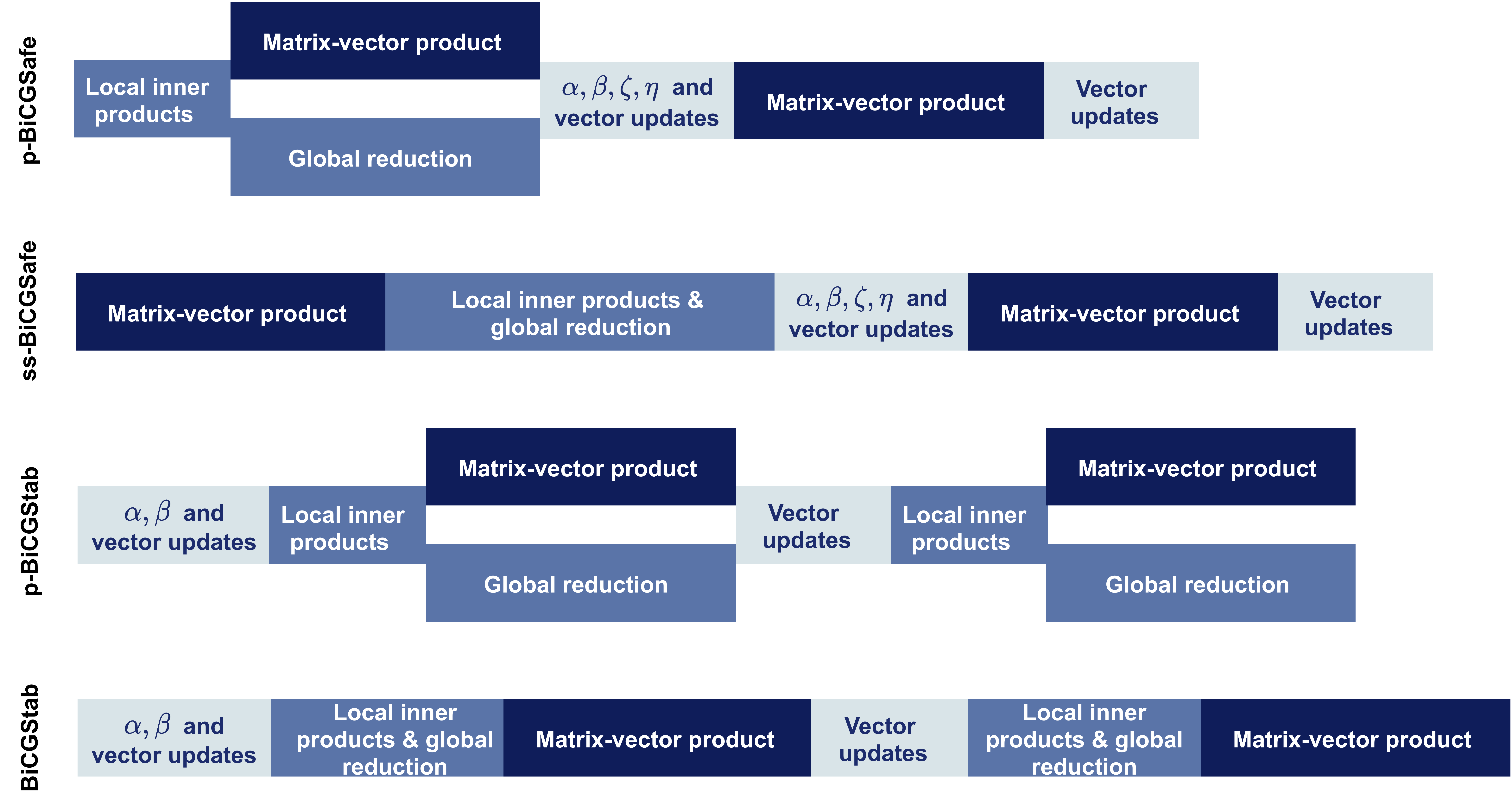}
\end{figure}

Because Alg. \ref{alg:PipelinedBiCGSafe} is generated by equivalent transformations of the expressions of the vector variables of Alg. \ref{alg:ssBiCGSafe2}, it is mathematically equivalent to Alg. \ref{alg:ssBiCGSafe2} in exact arithmetic.
However, as shown in Fig. \ref{fig:diagram}, the strategy of hiding communication costs of the inner product computation is applicable to Alg. \ref{alg:PipelinedBiCGSafe}, but not to Alg. \ref{alg:ssBiCGSafe2}. In Alg. \ref{alg:PipelinedBiCGSafe}, computation of the inner products (line $10$ and line $16$) can be performed simultaneously or in a manner that overlaps with the computation of matrix-vector multiplication $\bvec {As}_i$ (line $6$) because the value of $\bvec {As}_i$ is not necessary to calculate the values of inner products $a_i, b_i, ..., r_i$.

Figure \ref{fig:diagram} presents the difference between the two methods. Inner product computations that include a global reduction phase are overlapped with the matrix-vector multiplication computation in the pipelined BiCGSafe method. Actually, this overlap masks the communication latency of the inner product computations in the course of the matrix-vector computation. Figure \ref{fig:diagram} also shows a characteristic of the pipelined BiCGSafe and ssBiCGSafe that makes them superior to BiCGStab and pipelined BiCGStab: a single phase of global reduction by iteration. The performance estimate of the ssBiCGSafe method compared to iterative methods with two or three phases of global reduction is given in \cite{Fujjino2018}. The superiority of ssBiCGSafe is demonstrated with many matrices compiled from different analytic fields such as electromagnetic, thermal, structure, and fluid analysis. The pipelined BiCGSafe is constructed from ssBiCGSafe has inherited the property of a single phase of global reduction from ssBiCGSafe.

Table \ref{ta:computational_cost} presents the computational costs of the Pipelined BiCGSafe method in conjunction with conventional iterative methods of three types without preconditioning per iteration. In Table \ref{ta:computational_cost}, $nnz$ is the total number of non-zero elements. Also, $N$ is the number of rows in the matrix. Column $\textbf{\#} \mathbf{Ax}$ shows the number of floating point operations necessary to compute matrix-vector products. The star sign shows that one of the matrix-vector products is overlapped with a global reduction phase.
The diamond sign represents that both of matrix-vector products are overlapped with two global reduction phases \cite{Cools}. Column \textbf{\# memories} specifies the memory space to store the vectors used for the methods. Columns \textbf{\#} $\alpha \bvec x$ and  \textbf{\#} ($\bvec x + \bvec y$) show the number of floating-point operations for arithmetic operations in scalar multiplication and vector addition. Although more floating-point operations are used for the scalar multiplication and vector addition of the pipelined BiCGSafe method than for the ssBiCGSafe method, the overlap of matrix-vector product computation and global reduction communication can improve the overall computation time of the pipelined BiCGSafe method and engender a faster convergence than the standard ssBiCGSafe.

\newcolumntype{g}{>{\columncolor{Gray}}c}
\begin{table}[h]
\caption{Computational costs by one iteration of iterative methods of four kinds}
\label{ta:computational_cost}
\centering
\begin{tabular}{l g g g g g}
\toprule \rowcolor{white}
 &  \textbf{\#} $\bvec{ Ax}$ &  \textbf{\#} $\alpha \bvec x$ &  \textbf{\#} ($\bvec x + \bvec y$) &  \textbf{\#} $(\bvec x, \bvec y)$ & \textbf{\# memories} \\\rowcolor{white}
&\small{ ($\times nnz$)} & \small{($\times N$)} & \small{($\times N$)} & \small{($\times N$)} & \small{($\times N$)} \\
p-BiCGSafe & $2^{\star}$	 & 26  &  22 & 9 & 15  \\ \rowcolor{white}
ssBiCGSafe2 & 2	 & 16  & 14 &  9 & 10\\ 
p-BiCGStab & $2^{\diamond}$	 & 11  & 11 &  7  & 11 \\ \rowcolor{white}
BiCGStab & 2	 & 6  & 6 &  5 & 7 \\ 
\bottomrule
\end{tabular}
\end{table}

In the exact arithmetic, the pipelined BiCGSafe and ssBiCGSafe are mathematically equivalent, but in the finite precision arithmetic, the convergence characteristic of the pipelined BiCGSafe method can differ from that of the ssBiCGSafe because of rounding errors in arithmetic operations of floating-point numbers. Starting from the same initial value, the residue of the pipelined BiCGSafe can gradually differ from the ssBiCGSafe after each iteration.
In applications that require high accuracy, deviation of the residual vector can cause it to stagnate at a value greater than the required accuracy and can prevent the method from converging within the allowable number of iterations. To address this situation, we consider the residual substitution technique \cite{Cools, Vost1999} as described in the next section.

\section{Pipelined BiCGSafe Algorithm with Residual Replacement Technique}
To reduce rounding error effects,
we use the residual replacement technique \cite{Cools2018,Cools,Vost1999,Sleijpen1996} by setting the residual vector $\mathbf{r}_i$ and the auxiliary vectors  $\mathbf{q}_i, \mathbf{w}_i, \mathbf{v}_i, \mathbf{g}_i, \mathbf{s}_i$ back to their true values after each $m$ iteration as

\begin{align*}
\mathbf{r}_i &\gets \mathbf{b}_i - \mathbf{Ax}_i, \quad \quad
\mathbf{q}_i \gets \mathbf{Ao}_i,  \qquad \quad
\mathbf{w}_i \gets \mathbf{Au}_i,\\
\mathbf{l}_i &\gets \mathbf{At}_i,\qquad \quad \quad \ \ 
\mathbf{g}_i \gets \mathbf{Ay}_i,\qquad \quad \ 
\mathbf{s}_i \gets \mathbf{Ar}_i.
\end{align*}

The residual replacement should be avoided if $||\mathbf{r}_i||$ is small after a certain number of iterations $M$ because this replacement might impair the convergence of the iterative method.
Resetting the residual at its actual value reduces effects of rounding errors accumulated after each iteration and allows the pipelined BiCGSafe to converge to the same accuracy as the original BiCGSafe. Algorithm \ref{alg:PipelinedBiCGSafe-rr} shows the pipelined BiCGSafe method in conjunction with the residual replacement technique. We refer to it as p-BiCGSafe-rr for the simplicity. Values $m$ and $M$ (line $26$ and $38$), which indicate the epoch and maximum iterations to replace residuals, can be selected manually according to the estimated total number of iterations. The epoch value $m$ should be sufficiently large that the additional cost of the matrix-vector computations is negligible relative to the total costs of the method. Different selections of these values might lead to differences in the convergence behavior of the p-BiCGSafe-rr method.

\begin{algorithm}[H]
\caption{Pipelined BiCGSafe with residual replacement (p-BiCGSafe-rr)}
\label{alg:PipelinedBiCGSafe-rr}
\begin{algorithmic}[1]
\STATE Let $\bvec{x}_0$ be an initial guess; 
\STATE Compute $\bvec{Ax}_0$; \COMMENT{matrix-vector product}
\STATE $\bvec{r}_0 \gets \bvec{b}- \bvec{Ax}_0$;

\STATE Choose $\bvec{r}_0^*$ such that $(\bvec{r}_0^*,\bvec{r}_{0}) \neq 0$, e.g., $\bvec{r}_0^* = \bvec{r}_{0}$;

\FOR {$\textit{i = 0, 1...}$}

\STATE Compute $\bvec{As}_i$; \COMMENT{matrix-vector product}
\STATE Define $a_i \coloneqq (\bvec{s}_i,\bvec{s}_{i})$,
  $b_i \coloneqq (\bvec{y}_i,\bvec{y}_{i})$,
  $c_i \coloneqq (\bvec{s}_i,\bvec{y}_{i})$,
  $d_i \coloneqq (\bvec{s}_i,\bvec{r}_{i})$;
\STATE Define $e_i \coloneqq (\bvec{y}_i,\bvec{r}_{i})$,
  $f_i \coloneqq (\bvec{r}_0^*,\bvec{r}_{i})$,
  $g_i \coloneqq (\bvec{r}_0^*,\bvec{s}_{i})$,
  $h_i \coloneqq  (\bvec{r}_0^*,\bvec{t}_{i-1})$,
  $r_i \coloneqq  (\bvec{r}_i,\bvec{r}_i)$;
\IF {$i=0$} 
\STATE Compute $a_i, d_i, f_i, g_i, r_i$;\COMMENT{inner products}
\STATE $\beta_i \gets 0$; 
\STATE $\alpha_i \gets f_i/g_i$;
\STATE $\zeta_i \gets d_i/a_i$;
\STATE $\eta_i \gets 0$; 
\ELSE
\STATE Compute $a_i,b_i,c_i, d_i,e_i, f_i, g_i,h_i,r_i$; \COMMENT{inner products}
\STATE $\beta_i \gets (\alpha_{i-1}f_i)/(\zeta_{i-1}f_{i-1})$;
\STATE $\alpha_i \gets f_i/(g_i + \beta_i h_i )$;
\STATE $\zeta_i \gets (b_id_i -c_ie_i)/(a_ib_i- c_i^2) $;
\STATE $\eta_i  \gets (a_ie_i -c_id_i)/(a_ib_i- c_i^2) $;
\ENDIF

\STATE \textbf{if} $||\bvec{r}_i|| \leq \epsilon ||\bvec{r}_0||$ \textbf{stop}. \COMMENT{$||\bvec{r}_i||_2=\sqrt{r_i}$}
\STATE	$\bvec{p}_i \gets \bvec{r}_i +\beta_{i}(\bvec{p}_{i-1}-\bvec{u}_{i-1})$;
\STATE $\bvec{o}_i \gets \bvec{s}_i + \beta_{i}\bvec{t}_{i-1}$;
\STATE $\bvec{u}_i \gets \zeta_i \bvec{o}_i + \eta_i(\bvec{y}_{i} + \beta_{i}\bvec{u}_{i-1}) $;

\IF {$i\mod m = 0$ \hspace{.3em} \textbf{and}  \hspace{.3em} $i>0$
 \hspace{.3em} \textbf{and} $i<M$}
\STATE Compute $\bvec{Ao}_{i}$, $\bvec{Au}_{i}$; \COMMENT{matrix-vector products}
\STATE $\bvec{q}_{i} \gets \bvec{Ao}_{i}$; 
\STATE $\bvec{w}_{i} \gets \bvec{Au}_{i}$; 
\ELSE
\STATE $\bvec{q}_i \gets \bvec{As}_i + \beta_{i}\bvec{l}_{i-1}$; \COMMENT{$\bvec q_i \coloneqq \bvec {Ao}_i$ (Eqn. \ref{eq:q})}

\STATE $\bvec{w}_i \gets \zeta_i \bvec{q}_i + \eta_i(\bvec{g}_{i} + \beta_{i}\bvec{w}_{i-1}) $;
  \COMMENT{  $\bvec w_i \coloneqq \bvec {Au}_i$ (Eqn. \ref{eq:w})}
\ENDIF

\STATE $\bvec{t}_i \gets \bvec{o}_i - \bvec{w}_i $;
\STATE $\bvec{z}_i \gets \zeta_i\bvec{r}_i + \eta_i\bvec{z}_{i-1}-\alpha_i\bvec{u}_i $;

\STATE $\bvec{y}_{i+1} \gets \zeta_i\bvec{s}_i + \eta_i\bvec{y}_{i} - \alpha_i\bvec{w}_i $;

\STATE $\bvec{x}_{i+1} \gets \bvec{x}_{i}+\alpha_i \bvec{p}_i + \bvec{z}_i $;

\IF {$i\mod m = 0$ \hspace{.3em} \textbf{and}  \hspace{.3em} $i>0$ 
 \hspace{.3em} \textbf{and} $i<M$ } 
\STATE Compute $\bvec{Ax}_{i+1}$; \COMMENT{matrix-vector product}
\STATE $\bvec{r}_{i+1} \gets \bvec{b} - \bvec{Ax}_{i+1}$;
\STATE Compute $\bvec{At}_{i+1}$, $\bvec{Ay}_{i+1}$, $\bvec{Ar}_{i+1}$; \COMMENT{matrix-vector products}
\STATE $\bvec{l}_{i} \gets \bvec{At}_{i}$;
\STATE $\bvec{g}_{i+1} \gets \bvec{Ay}_{i+1}$;
\STATE $\bvec{s}_{i+1} \gets \bvec{Ar}_{i+1}$;
\ELSE
\STATE $\bvec{r}_{i+1} \gets \bvec{r}_{i} - \alpha_i \bvec{o}_i -\bvec{y}_{i+1}$;
\STATE Compute $\bvec{Aw}_i$; \COMMENT{matrix-vector product}
\STATE $\bvec{l}_i \gets \bvec{q}_i - \bvec{Aw}_i $; \COMMENT{  $\bvec l_i \coloneqq \bvec {At}_i$ (Eqn. \ref{eq:l}) }
\STATE $\bvec{g}_{i+1} \gets \zeta_i\bvec{As}_i + \eta_i\bvec{g}_{i} - \alpha_i\bvec{Aw}_i $;\COMMENT{  $\bvec g_{i+1} \coloneqq \bvec {Ay}_{i+1}$ (Eqn. \ref{eq:g}) }
\STATE $\bvec{s}_{i+1} \gets \bvec{s}_{i} - \alpha_i \bvec{q}_i -\bvec{g}_{i+1}$;\COMMENT{  $\bvec s_{i+1} \coloneqq \bvec {Ar}_{i+1}$ (Eqn. \ref{eq:sip1}) }
\ENDIF

\ENDFOR
\end{algorithmic}
\end{algorithm}

\section{Numerical Experiments}

In this section, we present numerical experiments and demonstrate the benefits obtained using our proposed methods. We first show numerical results of p-BiCGSafe in comparison with ssBiCGSafe2, BiCGStab and p-BiCGStab (pipelined BiCGStab \cite{Cools2018}) for several matrices from {\it SuiteSparse Matrix Collection} \footnote{\href {https://sparse.tamu.edu}{https://sparse.tamu.edu/}}. We then explain the advantage of the pipelined BiCGSafe method associated with residual replacement technique on some matrices that are difficult to solve. Lastly, we give a performance result of p-BiCGSafe on a supercomputer.\\
Computations are performed with double precision throughout this section. For all cases, the right-hand side vector of the linear system is set so that the solution is the unit vector. The initial solution $\mathbf{x}_0$ is set to the zero vector. Also, the initial residue $\mathbf{r}_0$ is set to a right-hand vector corresponding to the zero vector of the initial solution. 
The termination criterion for convergence is less than $\epsilon = 10^{-8}$ of the relative residual norm $||\mathbf{r}_i||_2/||\mathbf{r}_0||_2$.
The maximum number of iterations is set as $10^{4}$. Preconditioners are not used in all experiments. The use of preconditioners speeds up convergence, but it will mask the actual convergence behavior of the iterative methods.

Table \ref{ta:matrix_spec} presents specifications for test matrices from various analytic fields gathered from {\it SuiteSparse Matrix Collection}. In this table, $\mathbf{N}$ represents  the number of matrix rows, $\mathbf{nnz}$ denotes the total number of nonzero elements, $\mathbf{\kappa}$ denotes the 1-norm condition number, $\textbf{structure}$ indicates whether the matrix is symmetric or non-symmetric, and $\textbf{kind}$ indicates the field from which the matrix originates. The 1-norm condition numbers are estimated using the "condest" function provided with MATLAB software \footnote{\href {https://www.mathworks.com/}{https://www.mathworks.com/}}.

\begin{table}[ht]
\caption{Specifications of test matrices from the \textit{SuiteSparse Matrix Collection}}
\label{ta:matrix_spec}
\centering
\begin{tabular}{l g g g g g g g}
\toprule
\rowcolor{white}
&\textbf{N} &$\mathbf {nnz}$ &$\mathbf {nnz}$ \textbf{per row} & $\mathbf{\kappa}$ & \textbf{structure} & \textbf{kind}\\

atmosmodd& 1,270,432 & 8,814,880 & 6.9 &  $9.01 \times 10^{3}$  &non-sym. &fluid dynamics \\ \rowcolor{white}
poisson3Db& 85,623 & 2,374,949 & 27.7 & $1.65 \times 10^{5}$ &non-sym. & \ditto\\

sherman3& 5,005 & 20,033 & 4.0& $6.89 \times 10^{16}$ &non-sym.  & \ditto\\ \rowcolor{white} 

water\_tank& 60,740 & 2,035,281 & 33.5& $2.51 \times 10^{9}$  &non-sym. & \ditto \\

bcsstk18 & 11,948 & 149,090 &12.5& $6.48 \times 10^{11}$ &sym. & structural \\ \rowcolor{white} 

s3dkq4m2& 90,449 & 4,427,725 &49.0& $3.55 \times10^{11}$  &sym.  & \ditto \\ 

sme3Dc& 42,930 & 3,148,656 &73.0& $3.15 \times 10^8$ &non-sym. & \ditto \\ \rowcolor{white} 

epb3& 84,617 & 463,625 &5.5& $7.45 \times 10^4$ &non-sym. & \ditto \\

thermomech\_dK & 204,316 &2,846,228  &14.0& $2.18 \times 10^{19}$ &non-sym. & \ditto \\  \rowcolor{white}

tmt\_unsym& 917,825 &4,584,801 &5.0& $2.26 \times 10^{9}$ &non-sym. & electromagnetic \\ 

utm5940& 5,940 & 83,842 &14.1&  $1.90 \times 10^9$    &non-sym. & \ditto \\\rowcolor{white} 

xenon2& 157,464 & 3,866,688 &24.6& $1.76 \times 10^{5}$ &non-sym. & materials \\

\bottomrule
\end{tabular}
\end{table}

\subsection{Pipelined BiCGSafe on test matrices from \textit{SuiteSparse Matrix Collection}} 

\begin{figure}[!htb]
\caption{History of relative residual norms for test matrices.}
\label{fig:matrix_histories}
\includegraphics[width=0.48\textwidth]{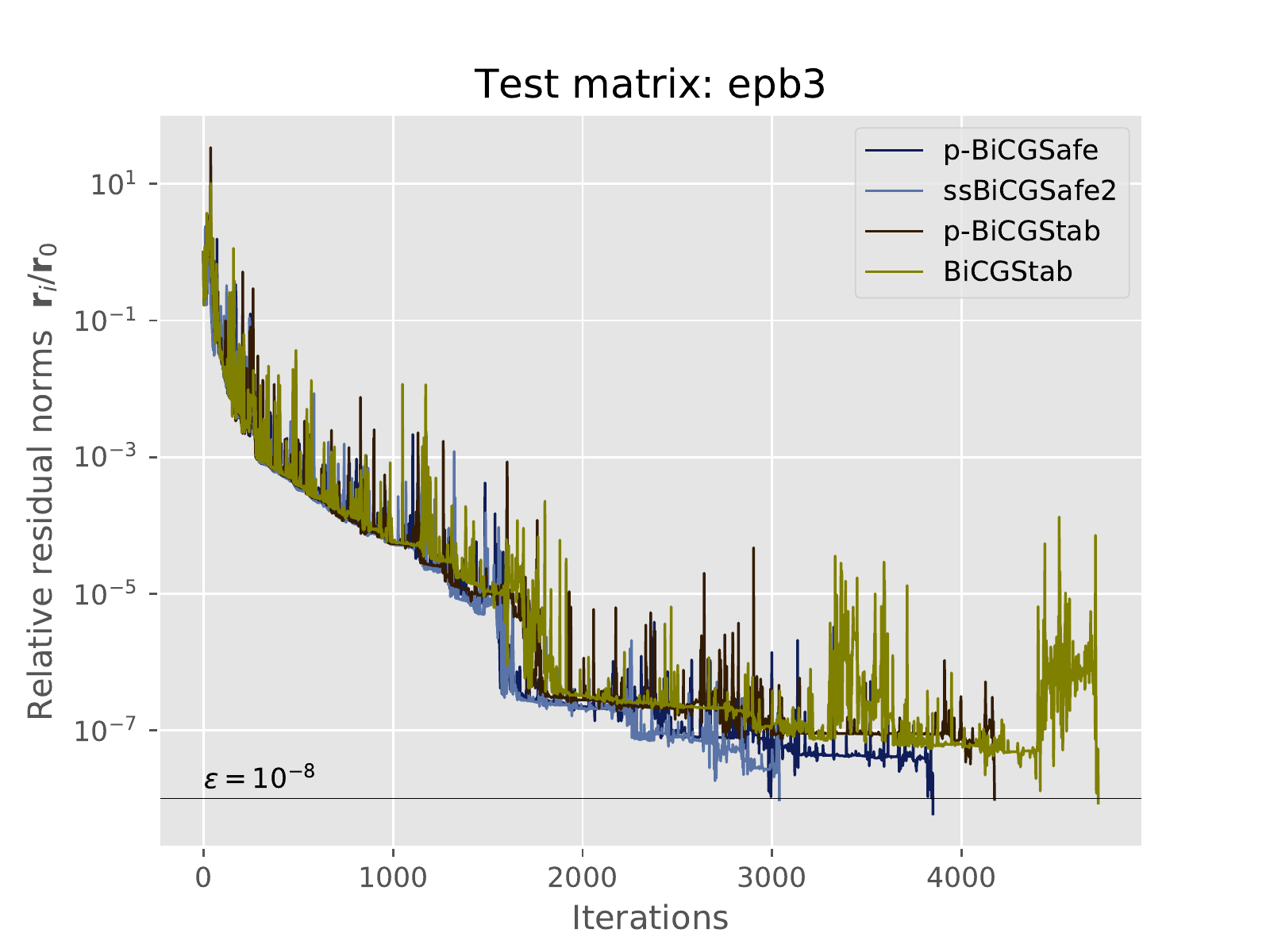}
\includegraphics[width=0.48\textwidth]{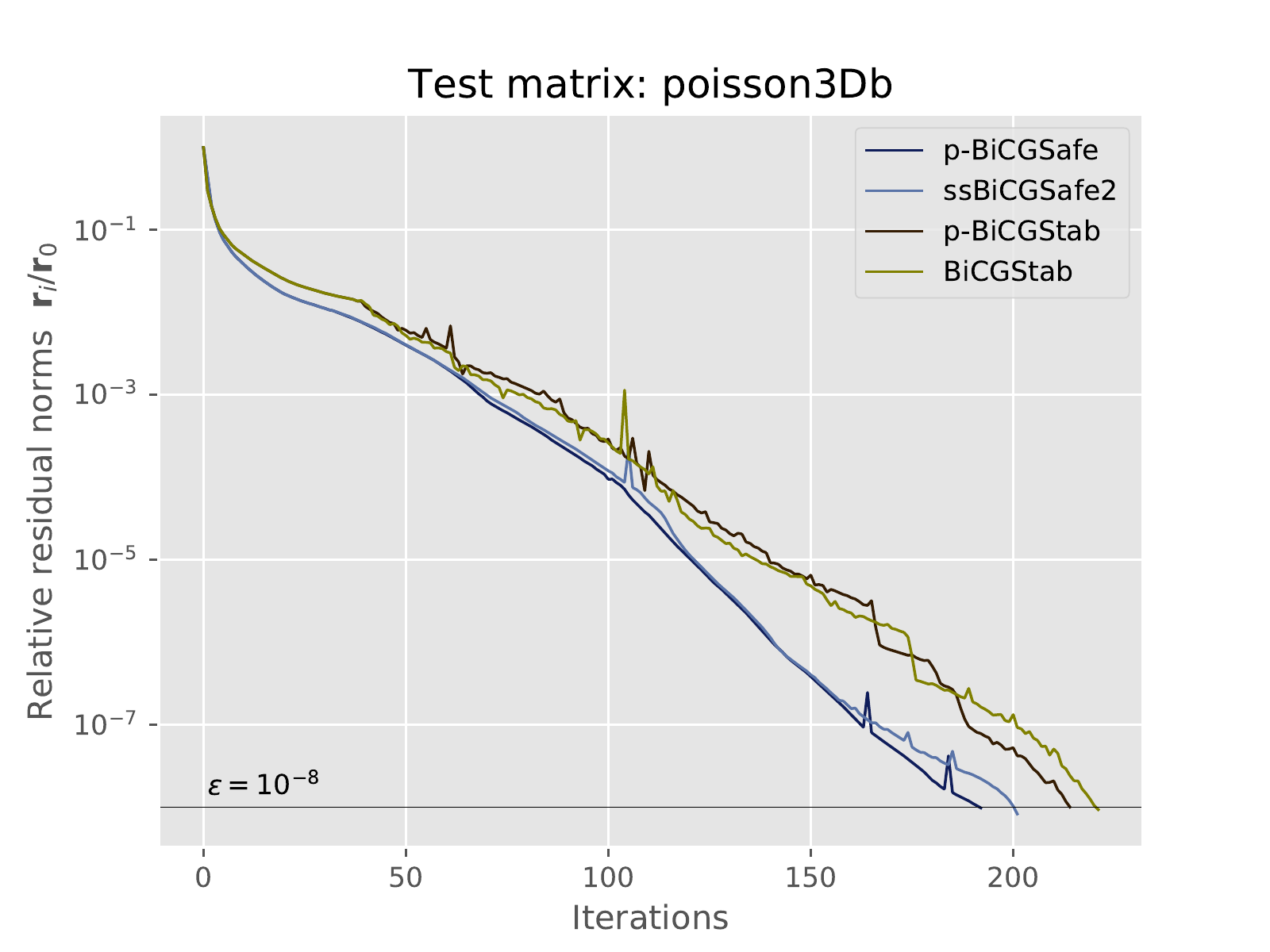}\\
\includegraphics[width=0.48\textwidth]{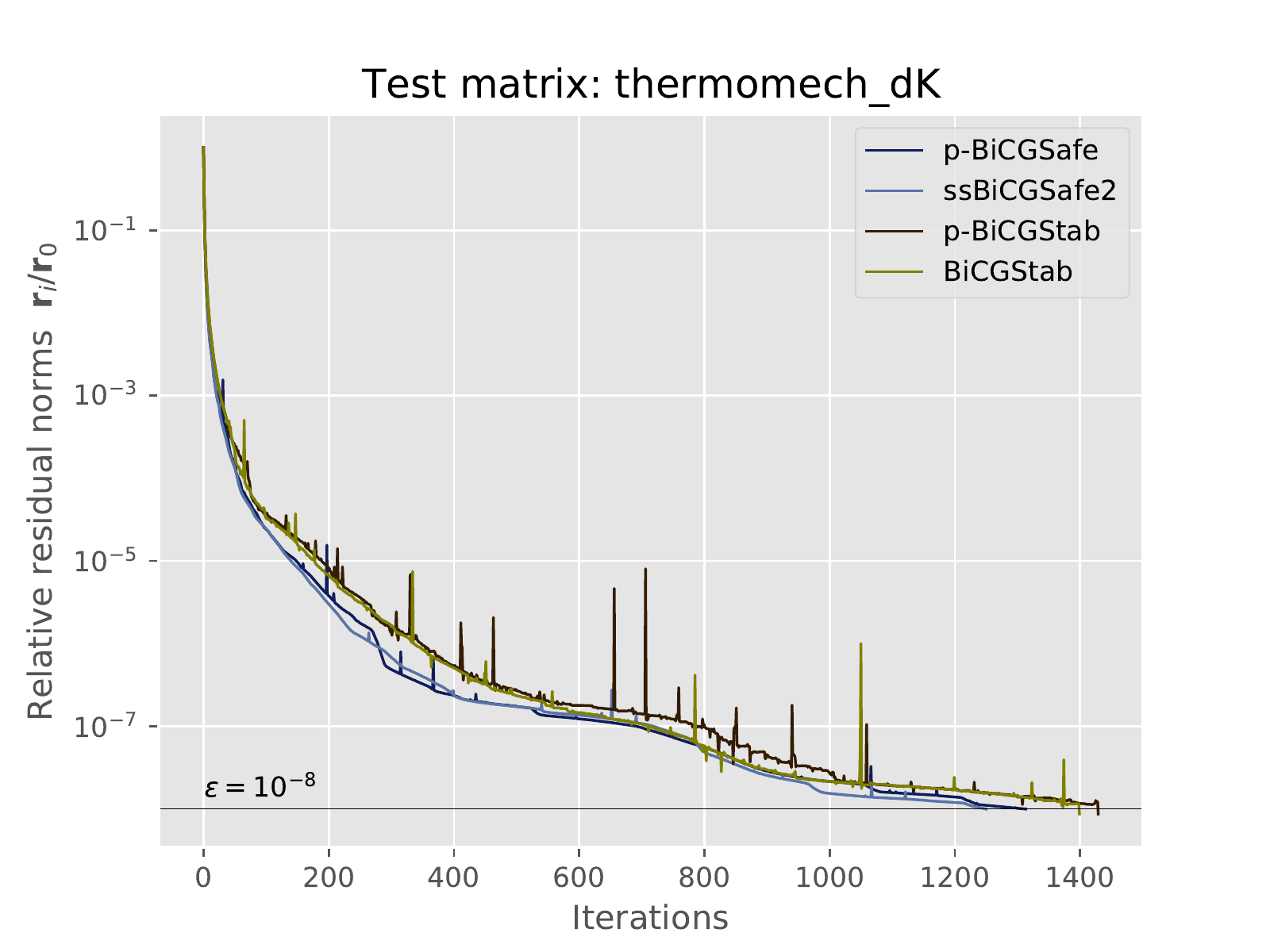}
\includegraphics[width=0.48\textwidth]{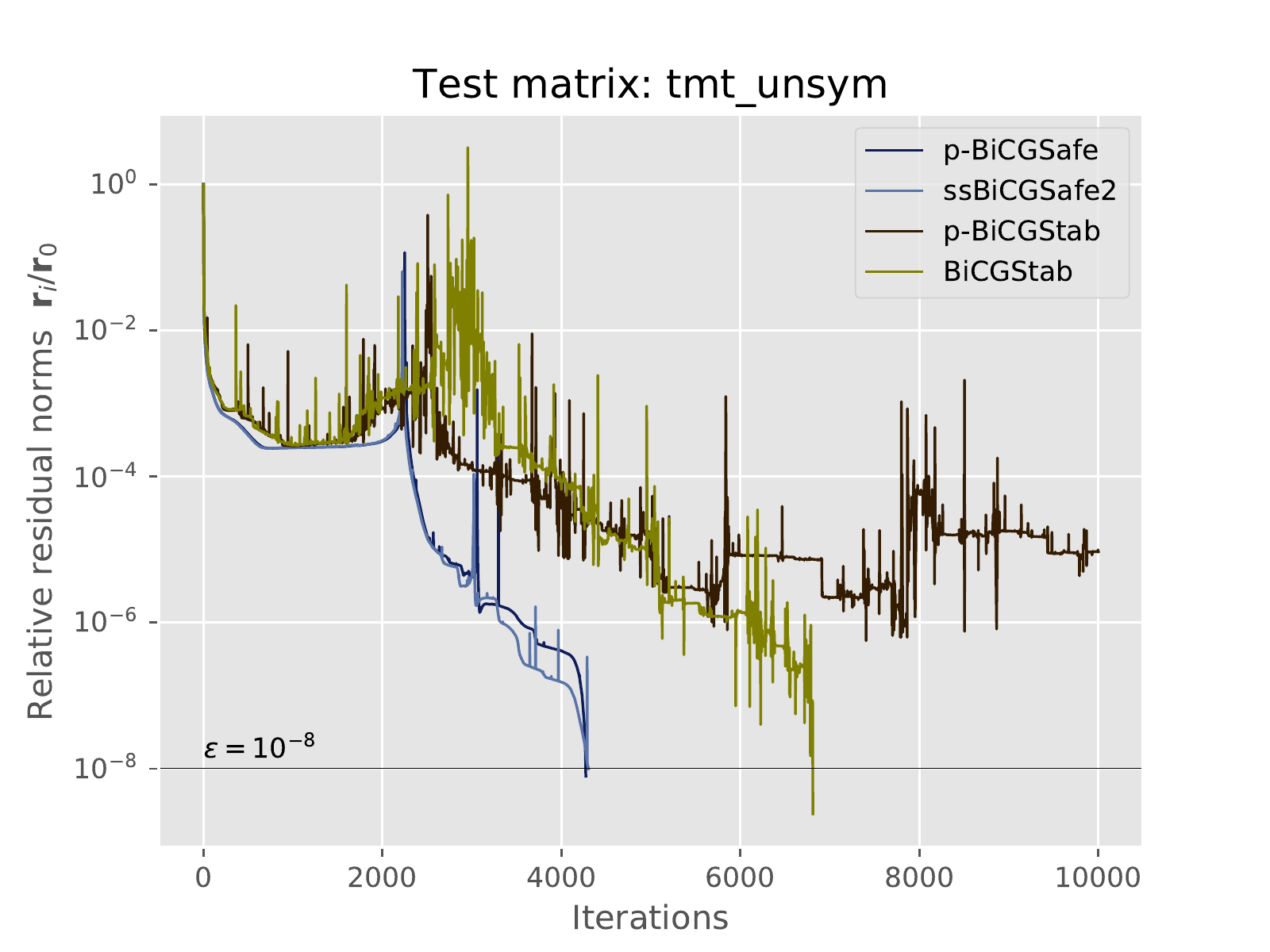}\\
\includegraphics[width=0.48\textwidth]{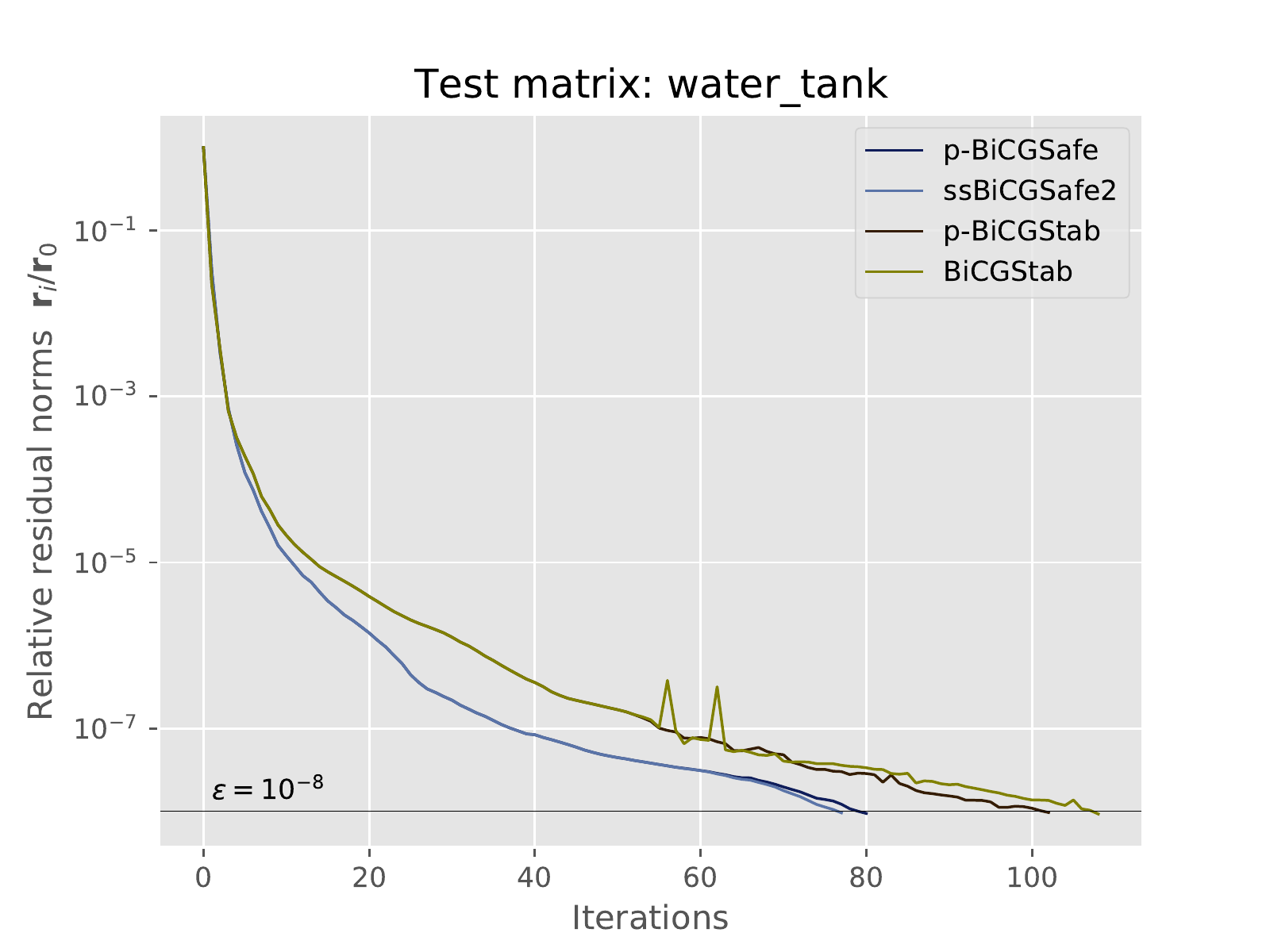}
\includegraphics[width=0.48\textwidth]{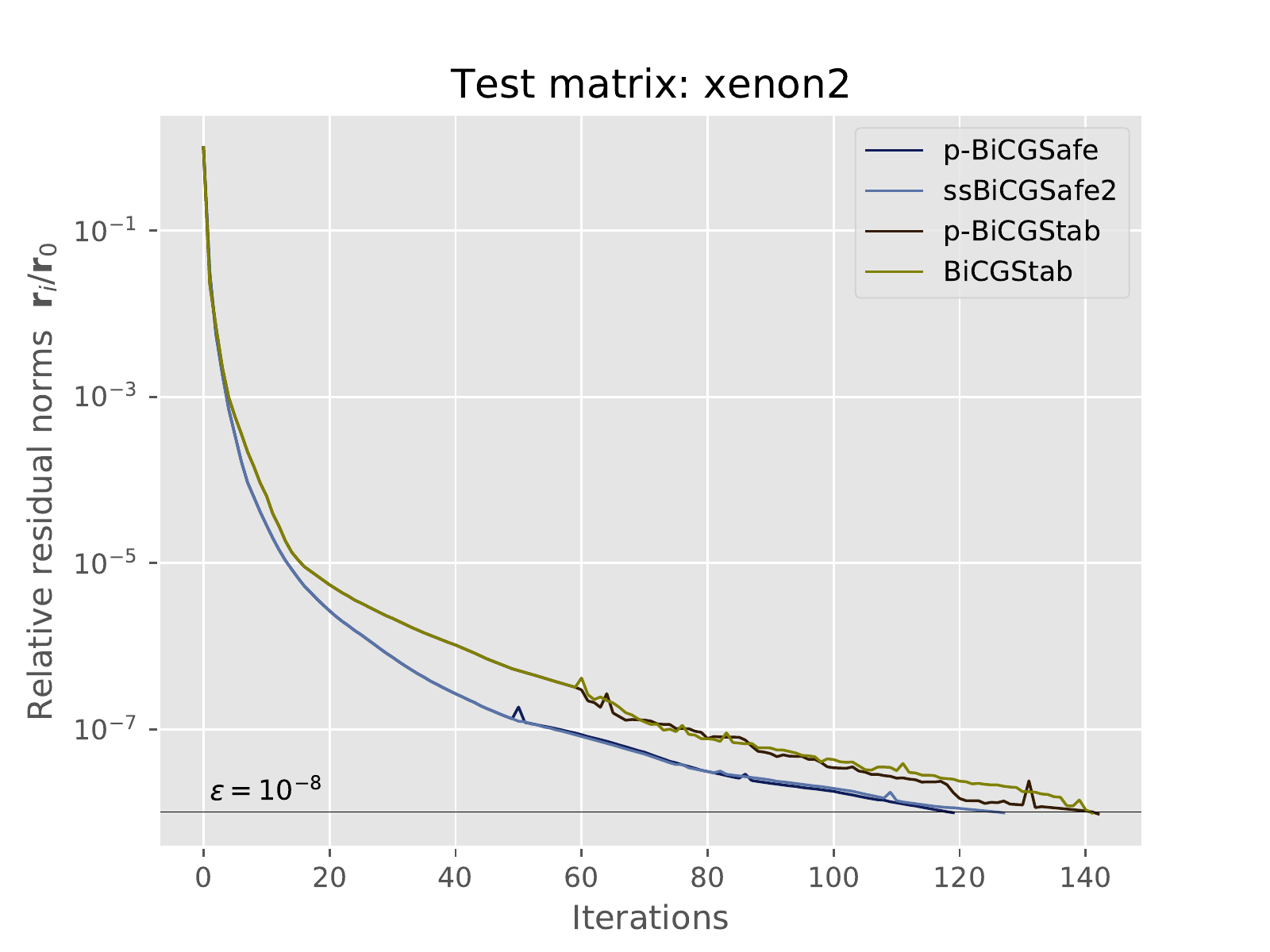}
\end{figure}

Figure \ref{fig:matrix_histories} presents histories of the relative residual norm for several test matrices chosen randomly for Table \ref{ta:matrix_spec} for the iterative methods of four types. In all sub-figures, both p-BiCGSafe and ssBiCGSafe2 graphs are nearly identical for the several dozen initial iterations. This strong similarity reflects the fact that both algorithms are equivalent in exact arithmetic terms, but they differ in finite precision arithmetic terms. 

The form of the graph of p-BiCGSafe and ssBiCGSafe2 appears smooth with fewer breaks than that of BiCGStab or p-BiCGStab. The graphs of p-BiCGSafe and ssBiCGSafe2 are shown below the graphs of BiCGStab and p-BiCGStab in all sub-figures: that is to say, the convergence features of the two tend to reach the successful termination criterion earlier than that of BiCGStab and p-BiCGStab. As shown in the sub-figure for the test matrices \textit{epb}, all four methods can achieve the successful termination criterion within the maximum iterations, but methods p-BiCGSafe and ssBiCGSafe2 are significantly earlier. Especially, with the test matrix \textit{tmt\_unsym}, the convergence behaviors of p-BiCGSafe and ssBiCGSafe2 are roughly twice as fast as that of p-BiCGStab. All these facts confirm the superior convergence feature of BiCGSafe compared to BiCGStab reported in \cite{Fujino2005}. These facts also reveal that p-BiCGSafe has inherited from ssBiCGSafe2 its outstanding convergence feature originating from BiCGSafe. The graphs in sub-figures of test matrices \textit{poisson3Db} shows that p-BiCGSafe achieved the desired accuracy sooner than ssBiCGSafe2 did. This random result exceeds expectations for the design of p-BiCGSafe, which has more arithmetic computations than ssBiCGSafe2 per iteration. This result does not imply that the convergence behavior of p-BiCGSafe is better than that of ssBiCGSafe2 in general.

\begin{table}[ht]
\caption{Number of iterations of iteration methods of four kinds on the test matrices}
\label{ta:num_its}
\begin{center}

\begin{tabular}{l g g g g  }
\toprule
\rowcolor{white}
&\textbf{p-BiCGSafe}& \textbf{ssBiCGSafe2}& \textbf{BiCGStab}& \textbf{p-BiCGStab}\\

atmosmodd&                      235&234   &235   &232 \\ \rowcolor{white}
poisson3Db&                     191&200   &220   &213   \\
sherman3&                       -&6400   &-   &-   \\ \rowcolor{white} 
water\_tank&                    79&76   &107   &101   \\  
bcsstk18&                       120&147   &141   &139   \\ \rowcolor{white} 
s3dkq4m2&                       628&598   &592   &585   \\ 
sme3Dc&                         918&817   &1170   &1053 \\ \rowcolor{white} 
epb3&                           3847&3037   &4719   &4712   \\
thermomech\_dK&                 1312&1249   &1398   &1428   \\  \rowcolor{white} 
tmt\_unsym&                     4272&4302   &6809   &-   \\ 
utm5940&                        -&5207   &-   &-   \\\rowcolor{white} 
xenon2&                         118&126   &140   &141   \\ 
\bottomrule
\end{tabular}
\end{center}
\end{table}

Table \ref{ta:num_its} shows the number of iterations to meet the stopping criterion of four iteration methods. The "-" symbol denotes that the method fails to converge. Also, ssBiCGSafe2 achieves safe convergence for all test matrices. p-BiCGSafe and BiCGStab have fewer convergence failures than p-BiCGStab. p-BiCGSafe provides better convergence stability than p-BiCGStab. The reason for this result might be readily apparent because BiCGSafe provides better convergence stability than BiCGStab, as shown in \cite{Fujino2005}.

\subsection{Pipelined BiCGSafe with residual replacement}

As described in the earlier section, algorithm p-BiCGSafe (Alg. \ref{alg:PipelinedBiCGSafe}) might fail to meet the convergence criterion for very challenging matrices that require numerous iterations to reach the convergence criterion because of accumulated round-off errors. As intended in the design goal of p-BiCGSafe-rr (Alg. \ref{alg:PipelinedBiCGSafe-rr}), we found two cases in the test matrices in which p-BiCGSafe-rr helps to improve the convergence behavior over p-BiCGSafe. \\

\begin{figure}[hbt!]
\caption{Convergence behavior improvement through residual replacement technique.}
\label{fig:Improved_convergence_behavior}
\includegraphics[width=0.48\textwidth]{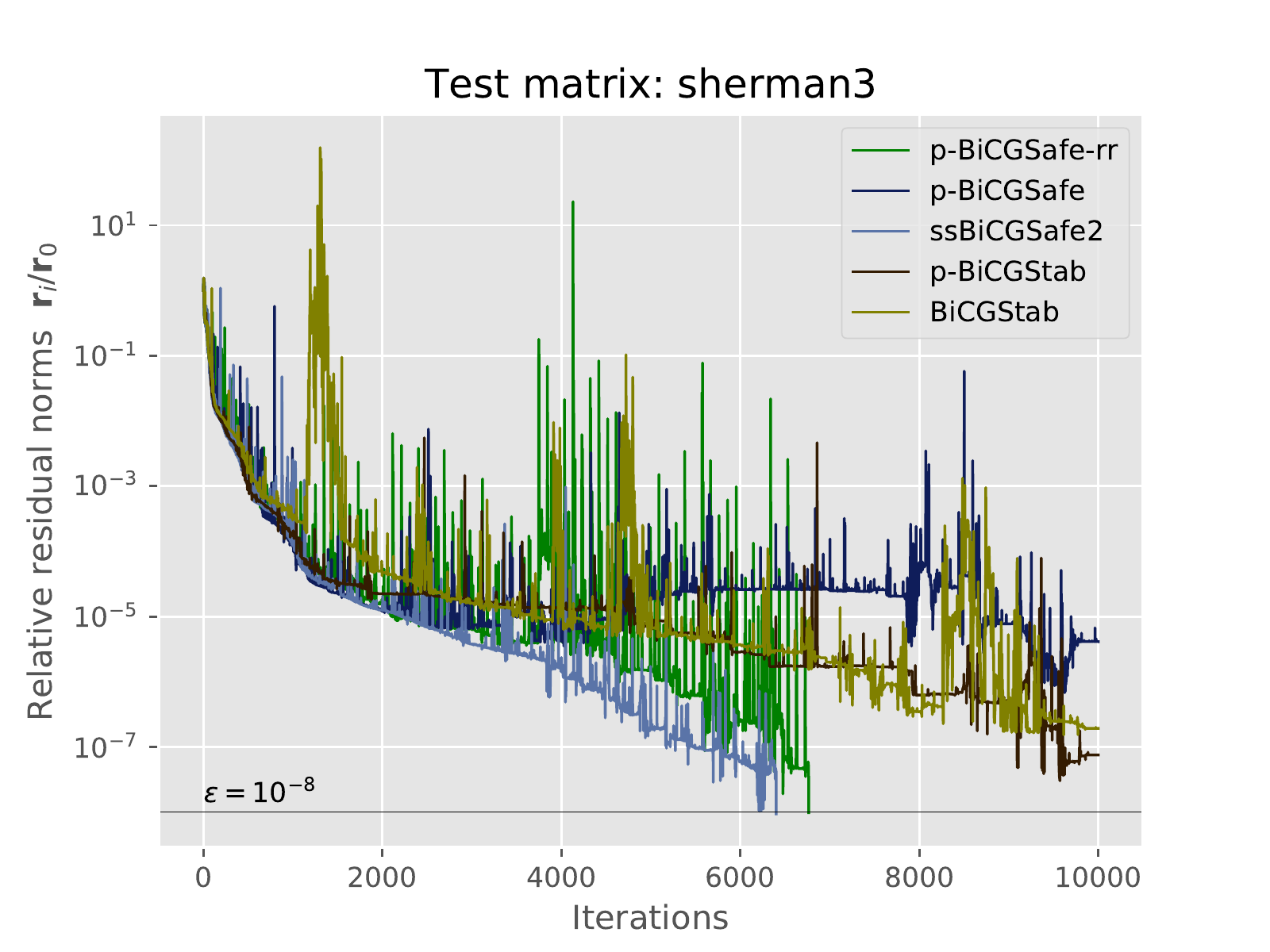}
\includegraphics[width=0.48\textwidth]{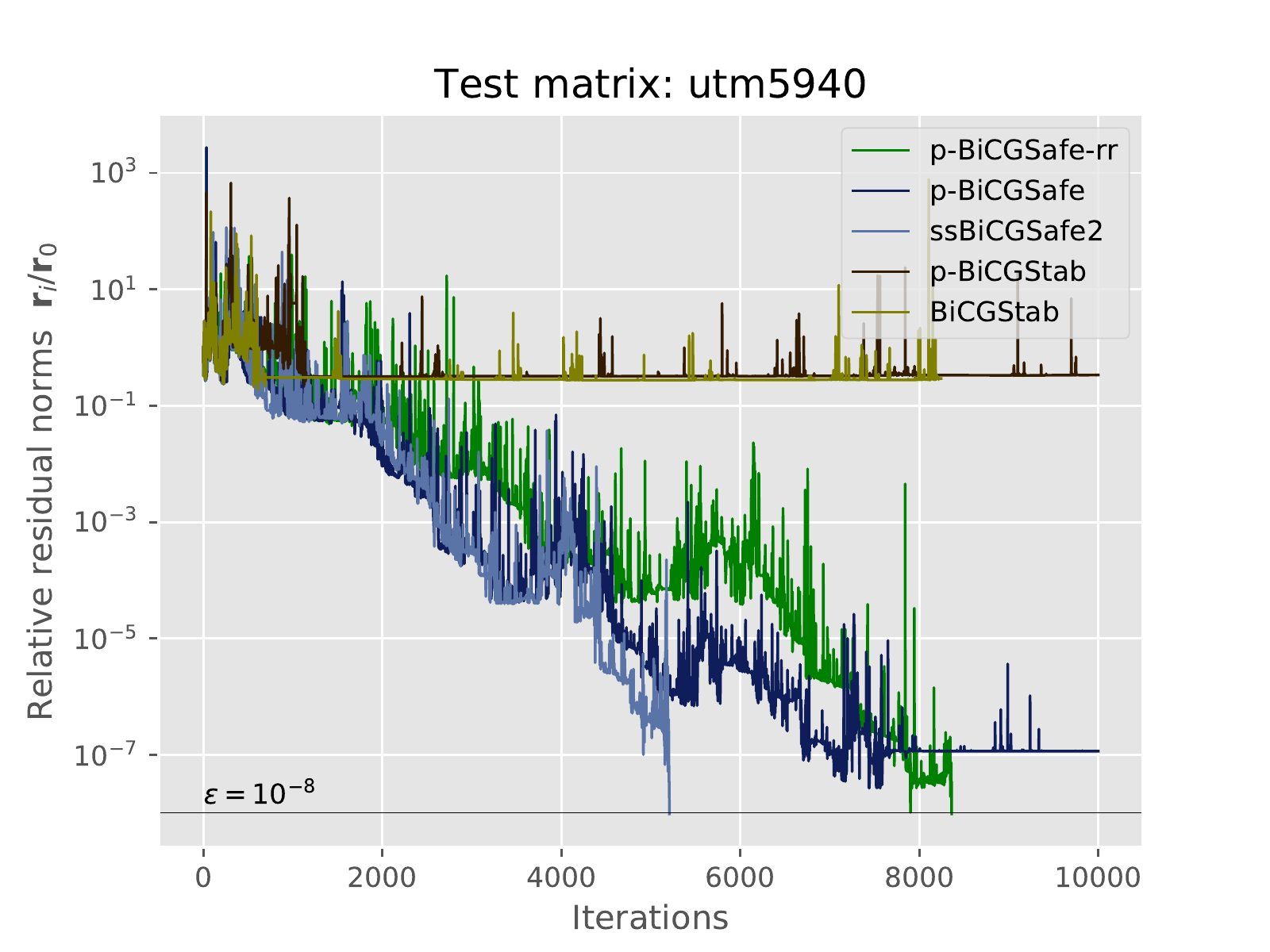}
\end{figure}

For test matrices \textit{utm5940} and \textit{sherman3}, Table \ref{ta:num_its} shows that the methods BiCGStab, p-BiCGStab, and p-BiCGSafe cannot achieve the termination criterion, whereas ssBiCGSafe2 can meet it within the maximum iteration. We applied p-BiCGSafe-rr for these two test matrices. The default epoch value of the residual replacement configured for this experiment is $m=100$. Also, p-BiCGSafe-rr achieves convergence for test matrix \textit{utm5940}, but not for test matrix \textit{sherman3}.
After changing the epoch value $m$ slightly to $m=96$, p-BiCGSafe-rr reaches convergence for this test matrix. Similarly to Fig. \ref{fig:matrix_histories}, Fig. \ref{fig:Improved_convergence_behavior} shows the relative residual norm histories for iterative methods of five types used to solve two test matrices \textit{sherman3} and \textit{utm5940}. Results show that the number of iterations of p-BiCGSafe-rr is greater than that of ssBiCGSafe2. This outcome can be viewed as a delayed convergence phenomenon that typically occurs when the residual replacement procedure changed the convergence behavior of ssBiCGSafe2. For this reason, p-BiCGSafe-rr should be regarded as a means of achieving the convergence when p-BiCGSafe is unable to do so. It should not be used as a complete replacement for p-BiCGSafe.

\subsection{Performance on a memory-distributed computing system}
   This section presents results showing that the pipelined BiCGSafe method can achieve results in less computation time on a memory-distributed HPC system than ssBiCGSafe2 can.
   
   We use the open-source PETSc library \footnote{\href {https://www.mcs.anl.gov/petsc/}{https://www.mcs.anl.gov/petsc/}} ver. 3.15 to implement the p-BiCGSafe and sBiCGSafe methods. MPI library MPICH \footnote{\href {https://www.mpich.org/}{https://www.mpich.org/}} ver. 3.4.2 was used to build the PETSc library from its source code.

\begin{figure}[hb!]
\begin{center} 
\caption{Computation times of two iterative methods versus their numbers of nodes.}
\label{fig:parallel_perform}
\includegraphics[width=0.65\textwidth]{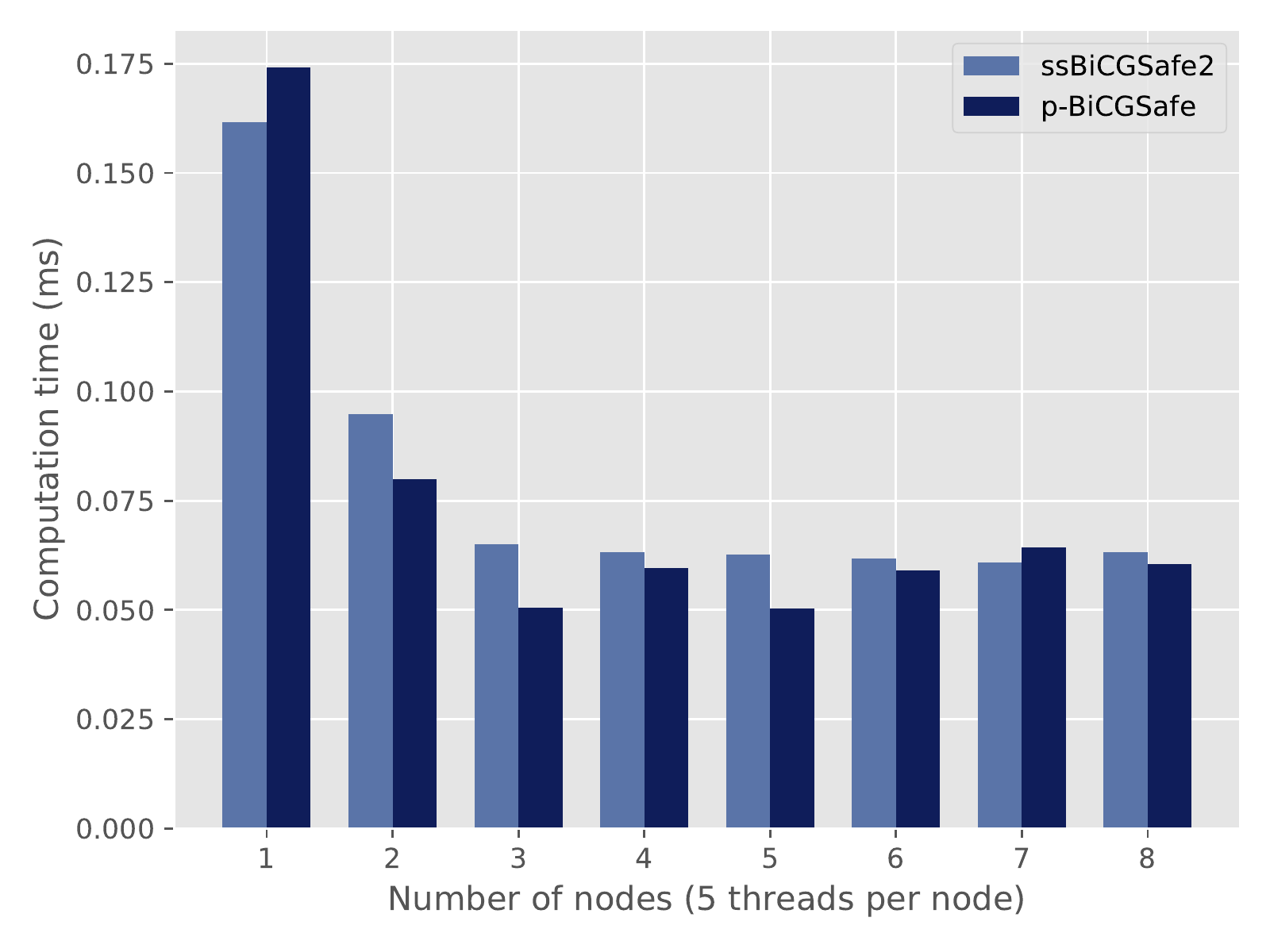}
\end{center}
\end{figure}

 The test matrix used for this experiment is \textit{Poisson3Db} from Table \ref{ta:matrix_spec}. We run the code of p-BiCGSafe and ssBiCGSafe2 on a cluster of eight nodes of a 2.40 GHz central processing unit (CPU, Xeon Gold6148; Intel Corp.). The MPI environment variables "MPICH\_ASYNC\_PROGRESS = 1" and "MPICH\_MAX\_THREAD\_SAFETY = multiple" are set so that communication and computation in MPI processes overlap without blocking. The function which computes the inner products in the program takes advantage of this setting. Figure \ref{fig:parallel_perform} shows the computation times of ssBiCGSafe2 and p-BiCGSafe methods. For a range of the number of nodes from 2 to 5, p-BiCGSafe is faster than standard ssBiCGSafe2. The computing time cannot be improved very much if the number of nodes exceeds 6. p-BiCGSafe reaches the minimum computation time when the number of nodes is 5. The effectiveness of pipelined iterative methods over its standard version has been well investigated in the works of Cools and Ghysels \cite{Cools, Ghysels2013,Ghysels2014}. Similarly, this experiment shows that the pipelined BiCGSafe method can be more efficient than the standard ssBiCGSafe2 method. \\
 As the number of nodes to compute increases, the time it takes to calculate the inner products increases. At some point, it becomes dominant and accounts for most of the total execution time of the algorithm \cite{Zhu2014}. Given a sparse matrix, a range of numbers of compute nodes exists for which total communication time of the inner products is covered completely by the matrix-vector computation time. In this case, the p-BiCGSafe method is faster than the ssBiCGSafe2 method. Also, a range of nodes exists where the p-BiCGSafe method cannot be faster than the ssBiCGSafe2 method. A comprehensive investigation of the behavior and performance of the proposed methods on HPC systems is beyond the scope of this report. It is left as a subject for future work.
 
 \section{Conclusions}
 We proposed communication of hiding pipelined iterative methods for solving large-scale linear systems on HPC systems. The proposed methods, developed based on the pipelined BiCGStab and single-synchronized BiCGSafe methods, inherited two outstanding properties from each method: hiding communication latency and using only one phase for global reduction per iteration. They combine the best of both worlds and therefore present more benefits for use on memory-distributed HPC systems. We evaluated the proposed methods using several numerical experiments. Compared to p-BiCGStab method, p-BiCGSafe method has a greater convergence characteristic; compared to ssBiCGSafe2 method, p-BiCGSafe method is more efficient on HPC systems in terms of computation time. As future work, we plan to apply the proposed method to a numerical simulation of large-scale fluid flow and structure systems of actual problems with supercomputers.

\section{Acknowledgments}
The lead author would like to thank Professor Emeritus Seiji Fujino at Kyushu University for his helpful comments on the research. This work was supported by JST CREST Grant Number JPMJCR15D1, Japan. Numerical simulations were partially carried out on the supercomputer system AFI-NITY at the Advanced Fluid Information Research Center, Institute of Fluid Science, Tohoku University.

\nocite{*}
\bibliographystyle{unsrt}
\bibliography{main}  
\end{document}